\AtBeginDocument{%
  }

\documentclass[acmsmall]{acmart}
\setcopyright{cc}
\setcctype{by-nc}
\acmJournal{PACMHCI}
\acmYear{2025} \acmVolume{9} \acmNumber{7} \acmArticle{CSCW504} \acmMonth{11} \acmDOI{10.1145/3757685}

\usepackage{placeins}
\begin{document}

\title{Two Sides to Every Story: Exploring Hybrid Design Teams’ Perceptions of Psychological Safety on Slack}



 \author{Marjan Naghshbandi}
\affiliation{%
  \institution{University of Toronto}
  \streetaddress{55 St George St}
  \city{Toronto}
  \state{Ontario}
  \country{Canada}
  \postcode{M5S 0C9}}
\email{marjan.naghshbandi@mail.utoronto.ca}

 \author{Sharon Ferguson}
\affiliation{%
  \institution{University of Waterloo}
  \streetaddress{xx}
  \city{Waterloo}
  \state{Ontario}
  \country{Canada}
  \postcode{xx}}
\email{sharon.ferguson@uwaterloo.ca}

 \author{Alison Olechowski}
\affiliation{%
  \institution{University of Toronto}
  \streetaddress{55 St George St}
  \city{Toronto}
  \state{Ontario}
  \country{Canada}
  \postcode{M5S 0C9}}
\email{a.olechowski@utoronto.ca}

\renewcommand{\shortauthors}{Marjan Naghshbandi, Sharon Ferguson, and Alison Olechowski}

\begin{abstract}
While the unique challenges of hybrid work can compromise collaboration and team dynamics, hybrid teams can thrive with well-informed strategies and tools that nurture interpersonal engagements. To inform future supports, we pursue a mixed-methods study of hybrid engineering design capstone teams' Psychological Safety (PS) (i.e., their climate of interpersonal risk-taking and mutual respect) to understand how the construct manifests in teams engaged in innovation. Using interviews, we study six teams' perceptions of PS indicators and how they present differently on Slack (when compared to in-person interactions). We then leverage the interview insights to design Slack-based PS indicators. We present five broad facets of PS in hybrid teams, four perceived differences of PS on Slack compared to in-person, and 15 Slack-based, PS indicators---the groundwork for future automated PS measurement on instant-messaging platforms. These insights produce three design implications and illustrative design examples for ways instant-messaging platforms can support Psychologically Safe hybrid teams, and best practices for hybrid teams to support interpersonal risk-taking and build mutual respect.
\end{abstract}

\ccsdesc[500]{Human-centered computing~Empirical studies in collaborative and social computing}

\keywords{Collaboration, Computer-Mediated Communication, Instant-Messaging Platforms, Psychological Safety, Team Dynamics, Teams}

\received{October 2024}
\received[revised]{April 2025}
\received[accepted]{August 2025}

\begin{teaserfigure}
\includegraphics[width=0.90\textwidth]{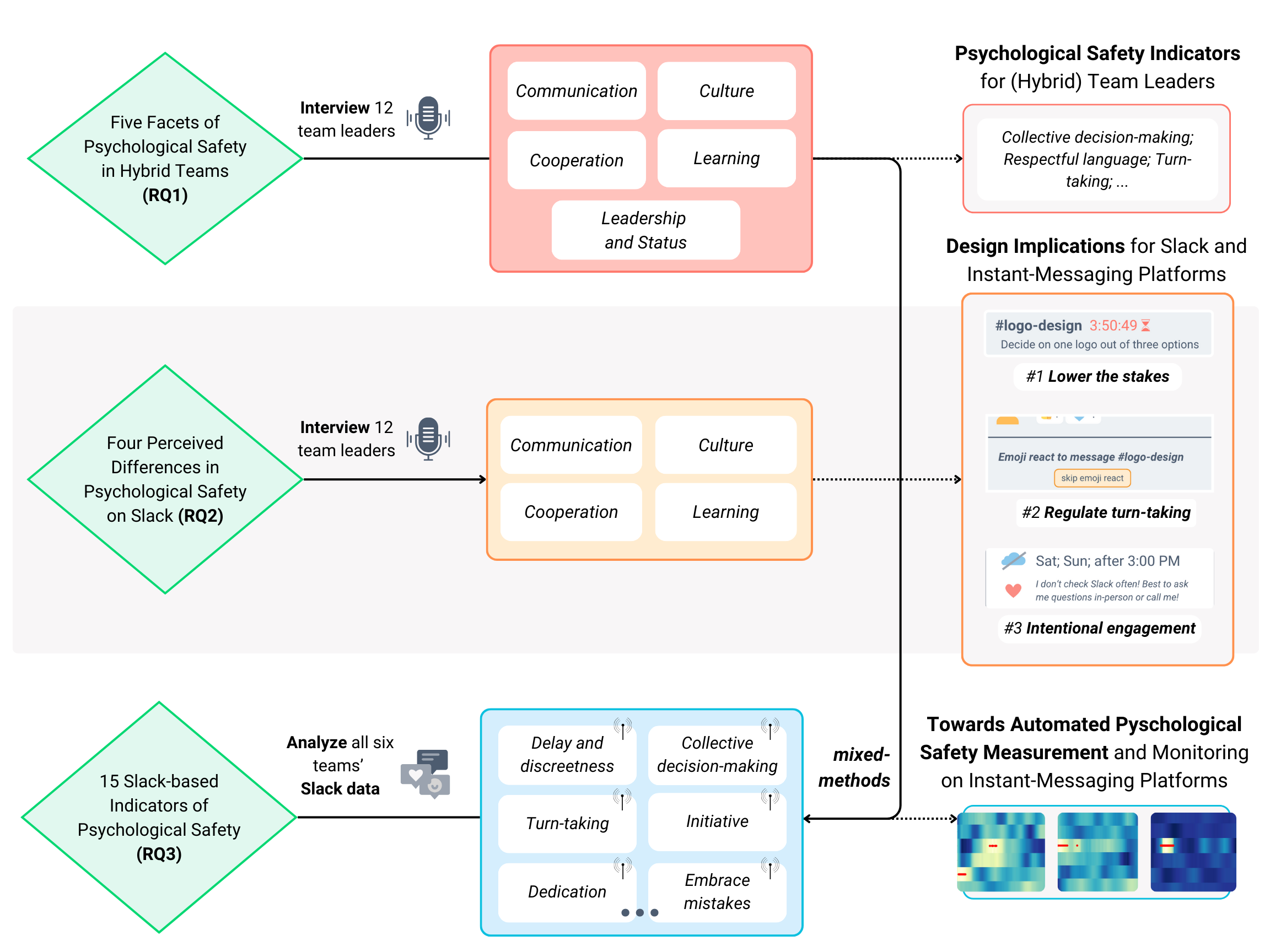}
\Description{This study asks  1. what are indicators of PS in hybrid teams?  2. how do hybrid teams perceive Psychological Safety (PS) to differ on Slack? To answer these two research questions, the study uses interviews with 12 hybrid team leaders, and quantitative analyses of six teams' Slack data. The results are organized by five facets of PS: Communication, Culture, Cooperation, Learning, and Leadership and Status. In response to RQ2, hybrid team leaders perceive Slack to affect the first four of the five PS facets. We use RQ2 findings to derive three design implications and produce three accompanying design examples for text-based ECPs to support PS. The design implications are called 1. lower the stakes, 2. regulate turn-taking, and 3. intentional engagement. Using the results from RQ1, we derive 15 Slack-based indicators of PS, helping us move towards an automated measure of PS online. }
\caption{Using thematic analysis of interview data and quantitative analyses of Slack data, we present \textbf{five facets of Psychological Safety (PS) in hybrid teams} for RQ1, \textbf{four perceived differences in PS on Slack compared to in-person} for RQ2, and \textbf{15 Slack-based indicators of PS} for RQ3. RQ1, RQ2, and RQ3 thus result in PS indicators for hybrid teams, design implications (incl. design examples) to support PS in hybrid teams' Slack communication, and preliminary, automated measures of PS, respectively.}
\end{teaserfigure}
\maketitle
\section{Introduction}
Future of Work research shows workers’ increasing preferences for hybrid work \cite{Teevan_Baym_Butler_Hecht_Jaffe_Nowak_Sellen_Yang_Ash_Awori_et_al._2022,Wigert_Harter_Agrawal_2023}. This may be explained by hybrid work’s positive effect on productivity, well-being, and work-life balance \cite{Angelici_Profeta_2024}, and hybrid communication's positive effects on job performance \cite{Zhang_Venkatesh_2013}. Still, hybrid work is not without its strains on teams, who endure the challenges of remote and in-person teams, as well as misalignment and unequal information access between teammates \cite{Duckert_Barkhuus_Bjørn_2023}. This mode of work can also threaten team relationships and identity, and contribute to burnout and communication obstacles 
\cite{Konovalova_Petrenko_Aghgashyan_2022}. 

Given these challenges, hybrid teams should prioritize relationship-building 
\cite{Bjorn_Busboom_Duckert_Bødker_Shklovski_Hoggan_Dunn_Mu_Barkhuus_Boulus-Rødje_2024, Brown_Hill_Lorinkova_2021, Teevan_Baym_Butler_Hecht_Jaffe_Nowak_Sellen_Yang_Ash_Awori_et_al._2022} and the establishment of both interpersonal trust \cite{Teevan_Baym_Butler_Hecht_Jaffe_Nowak_Sellen_Yang_Ash_Awori_et_al._2022} and a common ground \cite{Duckert_Barkhuus_Bjørn_2023,Teevan_Baym_Butler_Hecht_Jaffe_Nowak_Sellen_Yang_Ash_Awori_et_al._2022}. 
This motivates our focus on Psychological Safety (PS): a relationship-focused, team-level construct that is understudied \cite{Edmondson_Bransby_2023} yet critical \cite{Kirkman_Stoverink_2021} in hybrid teams. PS is defined as \textit{a team-wide belief that the team's climate is comprised of mutual respect, and safe for interpersonal risk-taking} \cite{Edmondson_1999}, signalled by teammates confidently engaging in behaviours that could trigger embarrassment, rejection, or punishment \cite{Edmondson_1999, Lechner_Tobias_Mortlock_2022}---such as asking questions, correcting others, and sharing knowledge \cite{O’Donovan_Van_Dun_McAuliffe_2020}. High PS teams have strong interpersonal trust and respect \cite{Nembhard_Edmondson_2006, Edmondson_1999}, encouraging authentic and inclusive climates \cite{Edmondson_Bransby_2023} where teammates are more satisfied and engaged \cite{Edmondson_Bransby_2023, Edmondson_Lei_2014, Newman_Donohue_Eva_2017}, and teamwork is more effective \cite{Edmondson_Bransby_2023}. PS is also critical for virtual teams' resilience \cite{Kirkman_Stoverink_2021}. PS promotes knowledge sharing in virtual communities \cite{Zhang_Fang_Wei_Chen_2010} and weakens virtuality's negative effect on team innovation \cite{Gibson_Gibbs_2006}. However, it is difficult for hybrid teams to leverage PS research findings in order to support their PS development. Existing literature primarily studies traditional, in-person teams, and even then, insufficiently explores specific team actions that affect PS \cite{Edmondson_Bransby_2023}. The minimal guidance, coupled with hybrid team leaders' only-partial visibility into team interactions, makes it challenging for these teams to monitor and nurture PS.


We investigate PS in hybrid teams to understand how the construct presents in this modern context,  
focusing on hybrid \textit{engineering design} teams to address their increasing collaborative information technology use \cite{marion2021transformation}. Our teams adhere to Lauring and Jonasson's \cite{lauring2025hybrid} recent definition of hybrid work: ``ongoing switches between interrelated traditional (analog/face-to-face) and non-traditional (digital/virtual) work modes in relation to modality, location, and temporality... '' \cite[p. 15]{lauring2025hybrid}. Specifically, they collaborate in-person and over Slack: an enterprise communication platform for instant-messaging (modality), from official workspaces and home (location), synchronously and asynchronously (temporality). Like other industry teams, engineering designers use collaborative tools for knowledge creation and information transfer  \cite{marion2021transformation}.
Further, this setting may broaden generalizability over healthcare delivery---a popular context in PS studies \cite{Edmondson_Bransby_2023}---where priority for patient safety is a key factor in PS development \cite{O’Donovan_Mcauliffe_2020_syst} but irrelevant to non-healthcare fields. Instead, engineers commonly work in teams \cite{Gidion_Buchal_2013} with the goal of innovation \cite{Petre}. Critical for engineering, PS encourages creativity, innovation, and performance \cite{Edmondson_Bransby_2023}, including teams’ tendency to satisfy customer needs and expectations \cite{Edmondson_1999}. Engineering design is also iterative \cite{Maier_Störrle_2011}, which necessitates learning from mistakes---PS conveniently encourages team learning behaviour 
\cite{Bresman_Zellmer-Bruhn_2013, Edmondson_1999, Harvey_Johnson_Roloff_Edmondson_2019}. Characteristics of engineering design teams, namely collaboration through knowledge sharing and creation, are presented as key success factors for knowledge work more broadly \cite{Sveiby_Simons_2002}, where workers commonly collaborate in a hybrid capacity \cite{Gartner_Newsroom_2023, The_Future_Forum_Team_2022}, highlighting our findings' applicability to many hybrid teams.

We pursue three research questions to investigate how PS is presented in hybrid engineering design teams and inform the design of best practices and interventions that optimize the health of their interpersonal engagements. First, we ask: \textbf{What are (positive and negative) PS indicators in hybrid teams?} (RQ1). To investigate the interplay between hybrid teams' dichotomous environment and PS, we then ask: \textbf{How (if at all) do hybrid teams perceive their PS to differ on Slack from in-person}? (RQ2). Following differences identified in RQ2, we note that team leaders may want to monitor PS online. Toward this goal, we ask: \textbf{To what extent are Slack-based PS indicators capable of differentiating hybrid teams' online PS trends?} (RQ3). We define PS indicators as \textit{actions that support PS and/or signals of high PS, as these actions support a cyclical relationship between supporting PS-building and also signalling its presence.}

We pursue a sequential exploratory design \cite{Creswell_2009} mixed-methods study with six hybrid teams engaged in engineering design capstone projects at a major U.S. institution. We use semi-structured interviews with team members who have high visibility into their team’s PS to answer RQ1 and RQ2. Our qualitative findings from RQ1 inform RQ3's quantitative analysis: we develop Slack-based PS indicators that are applied to instant-messaging data from teams' (public) Slack channels. With most PS studies depending on quantitative survey methods \cite{Edmondson_Bransby_2023, Newman_Donohue_Eva_2017}, our mixed-methods paper addresses the calls for qualitative insights on PS’ development and impact \cite{Frazier_Fainshmidt_Klinger_Pezeshkan_Vracheva_2017, Newman_Donohue_Eva_2017} and aligns with the praise of mixed-method approaches in PS studies \cite{Edmondson_1999, Edmondson_Lei_2014, O’Donovan_McAuliffe_2020_exploring}.


We identify five facets of PS: Communication, Cooperation, Culture, Learning, and Leadership and Status. RQ1 illustrates teams' perceptions of PS indicators across these five PS facets, and RQ2 reveals that four of these facets are perceived to be different on Slack compared to in-person---suggestive of \textit{a second side to PS}. Teammates are perceived to be more muted on Slack, causing miscommunications (Communication), less likely to help and trust each other on Slack (Cooperation), inclined to use Slack for ``\textit{mostly business}'' activities (Culture), and engaged in ineffective discussions and feedback-sharing over Slack (Learning). To monitor PS online, we present 15 Slack-based PS indicators---three for each PS facet. While only an initial step toward an automated measure, we show that these indicators can identify \textit{fluctuations in}, and \textit{differences between}, teams' PS on Slack over time.



This study \textbf{first} contributes to recent work on relational and communicative challenges in hybrid teams \cite{Duckert_Barkhuus_Bjørn_2023, Neumayr_Saatci_Rintel_Klokmose_Augstein_2022, Bjorn_Busboom_Duckert_Bødker_Shklovski_Hoggan_Dunn_Mu_Barkhuus_Boulus-Rødje_2024}---a context insufficiently studied in PS research \cite{Edmondson_Bransby_2023}---a five-facet PS framework that is grounded in Edmondson's \cite{Edmondson_1999} PS operationalization, but comprehensive and actionable in its makeup of five facets (containing 24 indicators). As such, we provide hybrid team leaders guidelines to help monitor PS through team Communication, Cooperation, Culture, Leadership and Status, and Learning. While PS' significance has been quantitatively explored in HCI \cite{Cao_Yang_Chen_Lee_Stone_Diarrassouba_Whiting_Bernstein_2021,Hastings_Jahanbakhsh_Karahalios_Marinov_Bailey_2018,Park_Santero_Kaneshiro_Lee_2021}, our qualitative findings unveil actionable insights to support teams and researchers.


\textbf{Second}, we contribute to recent literature on design opportunities in instant-messaging platforms' collaboration support \cite{Lee_Yuan_2024, chou2022because, chou2022did} the first understanding of perceived differences in the extent to which PS is effectively expressed on Slack compared to in-person across four PS facets, and we develop three design implications for Slack and related platforms' greater support of PS. \textbf{Third}, our Slack-based PS indicators in RQ3 extend HCI's use of analytics to study and support team phenomena \cite{Cao_Yang_Chen_Lee_Stone_Diarrassouba_Whiting_Bernstein_2021, Wang_Wang_Yu_Ashktorab_Tan_2022, Ferguson_Van_De_Zande_Olechowski_2024,Musick_Gilman_Duan_McNeese_Knijnenburg_O’Neill_2023}. In particular, we build on recent efforts to leverage instant-messaging data for this purpose \cite{Cao_Yang_Chen_Lee_Stone_Diarrassouba_Whiting_Bernstein_2021, Wang_Wang_Yu_Ashktorab_Tan_2022, Ferguson_Van_De_Zande_Olechowski_2024}, to inform the design of future automated online PS measurement on Slack and similar platforms. Future automated measures, grounded in in-depth empirical work, can support team leaders in monitoring various online interactions in hybrid teams.




\section{Related Work}
Here, we review the literature on remote/hybrid team challenges and PS, \textcolor{black}{clarifying our operationalization of PS.}

\subsection{Distributed Work (Remote and Hybrid) Challenges}
Distributed collaboration offers flexibility, but comes with potential harms such as worsened well-being \cite{Augstein_Neumayr_Schönböck_Kovacs_2023}. Its effects on productivity, coordination, knowledge sharing, trust, workload and commitment, also pose risks to collaboration and team relationships \cite{Breideband_Talkad_Sukumar_Mark_Caruso_D’Mello_Striegel_2022}. Since workers still have preferences for flexible work, researchers share technologies and best practices to support distributed collaboration \cite{Teevan_Baym_Butler_Hecht_Jaffe_Nowak_Sellen_Yang_Ash_Awori_et_al._2022}, including team dynamics and trust-building \cite{McVeigh-Schultz_Isbister_2022}. 

\subsubsection{Support for Hybrid Interactions}
Emerging in HCI is an emphasis on design that explicitly supports hybrid work, which we define as a mix of face-to-face and virtual, co-located and distributed, and (a)synchronous collaboration \cite{lauring2025hybrid}.
\citet{Bjorn_Busboom_Duckert_Bødker_Shklovski_Hoggan_Dunn_Mu_Barkhuus_Boulus-Rødje_2024} proposes design directions for hybrid work technology, including articulation work support, compensation for non-verbal asymmetries, and emphasis on inclusion. These design directions are informed by their findings that developing trust and instilling feelings of inclusion, among other relational work, is more emotionally laborious in hybrid teams, yet crucial for success \cite{Bjorn_Busboom_Duckert_Bødker_Shklovski_Hoggan_Dunn_Mu_Barkhuus_Boulus-Rødje_2024}. \citet{Liu_Van_Essen_Eggen_2024} presents a design space for improved informal communication and social awareness in hybrid work, which emphasizes social cues and activity as information carriers, and lightweight interactions (e.g., emoji responses). In this work, we study PS, one element of relational work, in hybrid teams, to inform PS-centric guidance and design opportunities for hybrid teams and their technology.

\subsubsection{Support for Instant-Messaging} Instant-messaging platforms are among the technologies used by hybrid teams for text-based communication \cite{Bjorn_Busboom_Duckert_Bødker_Shklovski_Hoggan_Dunn_Mu_Barkhuus_Boulus-Rødje_2024, neumayr2018domino}. As \citet{Lee_Yuan_2024} recently described, research on the impact of instant-messaging platforms---specifically those used by modern organizations (e.g., Slack, Microsoft Teams)---is limited.  \citet{Lee_Yuan_2024} shows that unique social interactions play out on these platforms, such as subordinates' self-censorship and impression management tendencies. With the need for more research on how social variables and work context interplay with instant-messaging platform use \cite{Lee_Yuan_2024, gupta2013should}, we study how hybrid teams perceive PS to differ on Slack, compared to their in-person interactions.


\subsection{Psychological Safety}
As the previous section discussed relational challenges in hybrid teams, our work studies these teams through the lens of PS---a construct that mitigates obstacles to effective teamwork \cite{Edmondson_Bransby_2023}, yet is insufficiently explored in teams of modern work arrangements \cite{Edmondson_Bransby_2023}. Below we discuss how we operationalize PS, its benefits, and its HCI research presence.

\subsubsection{Operationalization}
\textcolor{black}{As a widely studied construct, PS has amassed a number of descriptions. For instance, the active \cite{Edmondson_Lei_2014} team-level characterization of PS entails “a coherent interpersonal climate...[in which there exists] a blend of [interpersonal] trust, respect for each other's competence, and caring about each other as people'' \cite[p. 375]{Edmondson_1999}. The primary PS measure---Edmondson's \cite{Edmondson_1999} seven-item scale---describes PS as comprised of safety in admitting mistakes, raising tough issues, taking risks, asking for help, and the overarching feeling that one can be themselves, without fear of being rejected, having their efforts undermined, and skills devalued. These perceptions are often thought to be cyclical \cite{hsiang2019exploring}---feeling safe to admit mistakes is a signal of team PS, and continuing to admit these mistakes, or reacting positively when others do, will reinforce this perception among the team, supporting higher PS. Moreover, authors have yet to sufficiently explain how PS unfolds \cite{Edmondson_Lei_2014}, making it challenging to distinguish \textit{signals} of PS from \textit{actions that encourage} it. Thus, in this work, our detailed description of PS interpretations in hybrid teams may capture acts that signal PS and/or contribute to building greater PS---both of which likely exist in a positive feedback loop.}

\textcolor{black}{Since the popularization of PS, it is sometimes used as a term to encompass all actions that create a broadly positive team environment. As such, it can often be confused with related positive team constructs, such as trust or well-being. However, PS is its own, unique construct. \citet{Lechner_Tobias_Mortlock_2022} describes that trust exists between two people, and focuses on how secure a teammate is with \textit{another} teammate taking interpersonal risks, whereas PS ``permeates an entire group'' \cite[p. 1]{Lechner_Tobias_Mortlock_2022}, and concerns how secure teammates are with their \textit{own interpersonal risk-taking}, based on their perceptions of how the group will likely respond to their risks. While interpersonal trust likely enables PS, PS requires mutual respect among teammates, too \cite{Edmondson_1999, Edmondson_Lei_2014}. Meanwhile, worker well-being is a multi-faceted construct, comprised of social, subjective, and eudaimonic components \cite{Teevan_Baym_Butler_Hecht_Jaffe_Nowak_Sellen_Yang_Ash_Awori_et_al._2022}. While PS is associated with improved well-being \cite{loudoun2025critical, wang2022servant, erkutlu2016benevolent}, well-being's components, such as actualization (social), satisfaction (subjective), and sense of purpose (eudaimonic) \cite{fisher-2014}, fall outside of PS' conceptualization. Staying true to \citet{Edmondson_1999}, we operationalize PS as a team climate of \textit{interpersonal risk-taking and mutual respect}. This definition informs our data collection and analysis.}



\subsubsection{PS Findings}
Despite the increasingly nuanced findings about PS \cite{Edmondson_Bransby_2023}, its impact on individual-, team-, and organization-level outcomes remains widely recognized \cite{Edmondson_Lei_2014}, including innovation \cite{Gu_Wang_Wang_2013, Mura_Lettieri_Radaelli_Spiller_2016}, speaking up \cite{Bienefeld_Grote_2014, Edmondson_Bransby_2023, Edmondson_Lei_2014}, knowledge sharing  \cite{Liu_Keller_2021, Mura_Lettieri_Radaelli_Spiller_2016}, learning \cite{Bresman_Zellmer-Bruhn_2013, Creon_Schermuly_2019, Edmondson_1999, Guchait_Madera_Dawson_2015, Harvey_Johnson_Roloff_Edmondson_2019, Hassan_Jiang_2021, Liu_Hu_Li_Wang_Lin_2014, Ortega_Van_Den_Bossche_Sánchez-Manzanares_Rico_Gil_2013, Wilhelm_Richter_Semrau_2019}, and performance \cite{Edmondson_1999, Koopmann_Lanaj_Zhou_2014, Liu_Keller_2021, Malhotra_Ahire_Shang_2017, Ortega_Van_Den_Bossche_Sánchez-Manzanares_Rico_Gil_2013}. PS mitigates the negative effect of obstacles to effective teamwork \cite{Edmondson_Bransby_2023}, including perceived dividing lines among subgroups \cite{Chen_Wang_Zhou_Chen_Wu_2017}, functional dominance in cross-functional teams \cite{Malhotra_Ahire_Shang_2017}, employee status \cite{Bienefeld_Grote_2014}, age diversity \cite{Gerpott_Lehmann-Willenbrock_Wenzel_Voelpel_2021}, and virtuality \cite{Gu_Wang_Wang_2013}. The bulk of PS research has studied traditional, in-person teams, especially healthcare teams \cite{Edmondson_Bransby_2023}, where priority for patient safety is a key PS theme \cite{O’Donovan_Mcauliffe_2020_syst}. A notable contribution here is the set of PS categories for healthcare teams in \citet{O’Donovan_Van_Dun_McAuliffe_2020}. Ultimately, our focus on an under-explored work context in PS research (hybrid engineering design teams) motivates our reliance on Edmondson's \cite{Edmondson_1999} original operationalization (which is still widely used \cite{Edmondson_Bransby_2023, Edmondson_Lei_2014}), and inductive pursuit of PS indicators in this setting. 

\subsubsection{PS Findings in HCI}
Technology design and evaluation studies rarely evaluate effects on team-level social phenomena, including PS \cite{Harris_Gómez-Zará_DeChurch_Contractor_2019}. Still, the limited literature here does suggest PS' significance for positive outcomes. Quantitatively, \citet{Park_Santero_Kaneshiro_Lee_2021} shows PS to be indirectly associated with team efficacy; \citet{Cao_Yang_Chen_Lee_Stone_Diarrassouba_Whiting_Bernstein_2021} finds some human-coded PS labels in messages to be positive predictors of team viability; and, \citet{Hastings_Jahanbakhsh_Karahalios_Marinov_Bailey_2018} shows that PS perceptions are significantly related to subjective team experience measures in User Interface Design and Software Engineering student teams. With these findings being primarily quantitative, we study teams' perceptions of PS using interviews, to unveil actionable and rich insights on how it presents.

\citet{Ferguson_Van_De_Zande_Olechowski_2024} uses PS literature to inform an exploratory study of Slack behaviours for automated PS measurement, envisioning a future where managers can monitor team PS via digital tools and intervene as necessary. The findings \cite{Ferguson_Van_De_Zande_Olechowski_2024} show differences between two teams, including more equal contributions, targeted appreciation, direct feedback requests, and use of emoji reactions in the team with higher PS. Future work is recommended to use interviews and more teams to triangulate findings and advance automated PS measurement \cite{Ferguson_Van_De_Zande_Olechowski_2024}, which we do in this work.  

This section illustrates gaps in support for hybrid teams, including solutions for relationship-building and trust. As PS enables successful team outcomes, this work will inform future best practices, interventions, and collaborative technology design on how to encourage PS in hybrid teams. To this end, we investigate \textbf{PS indicators in hybrid teams (RQ1)}, \textbf{perceptions of how PS differs on Slack from in-person (RQ2)}, and \textbf{Slack-based PS indicators (RQ3)}. 

\section{Methods}
Here we describe our research setting, data collection, and analysis methods.

\subsection{Setting}
We study six teams from the 2023 run of an engineering design capstone course at a major U.S. institution. Working under the same access to resources (budget, facility access, etc.), expectations (Slack use, etc.), and 12-week timeline, teams design a high-fidelity alpha prototype of a novel consumer product that teams may proceed to commercialize once the semester is complete. 
Each team comprises 14--16 students, including two or three Project Managers (PMs) and two Human Resources Managers (HRMs). The PMs lead project management efforts, ensuring the final product meets technical specifications and schedule requirements. The HRMs monitor and regulate team dynamics, ensuring all teammates feel supported. Both roles also entail technical tasks like any other teammate.

\subsubsection{Demographics}\label{methods:demographics} We report the demographics of our consenting participants, which represent 91\% of all students enrolled in the course. Regarding gender demographics, 49\% identify as women, 46\% identify as men, 4\% prefer not to disclose, and 1\% identify as gender-fluid or non-binary. Additionally, 41\% identify as White, 28\% identify as Asian, 8\% identify as Hispanic or Latino, 4\% identify as Black or African American, and 1\% prefer not to disclose. The remaining 19\% of participants identify with two or more races/ethnicities (incl. American Indian or Alaskan Native, Asian, Black or African American, Hispanic or Latino, Native Hawaiian or Other Pacific Islander, and White). Gender and racial/ethnic diversity is maintained at the team-level, wherein 20\%--57\% of teammates identify as men and 25\%--57\% identify as White across the six teams. However, diversity is more varied among team leaders, who are the subjects of our interview study (described below): 20\%--80\% of PMs and HRMs identify as men and 17\%--80\% identify as White across the six teams.


\subsubsection{Our Hybrid Teams}
Our teams satisfy Lauring and Jonasson's \cite{lauring2025hybrid} definition of \textit{collaborative hybrid work}---they engage in face-to-face and virtual (hybrid \textit{modality}), co-located and distributed (hybrid \textit{location}), and synchronous and asynchronous collaboration (hybrid \textit{temporality}). Each team has dedicated three-hour in-person meetings each week, along with ad-hoc meetings in lab spaces as needed. In-person meetings are especially critical for physical prototyping activities. Teams are instructed to use Slack as their exclusive communication platform to reach teammates at different locations, hence each team is provided a Slack workspace for synchronous (e.g., live discussions) and asynchronous (e.g., offline updates) collaboration. The exclusive use of Slack additionally ensures instructors' visibility into team progress. Our second RQ studies how teams perceive PS on Slack to differ from in-person/face-to-face interactions---thus, we compare the only two settings that are systematically enforced by our hybrid teams' course guidelines and resources.
\subsection{Mixed-Methods Approach} We pursue a mixed-methods study of PS, as recommended by past work \cite{Edmondson_1999, Edmondson_Lei_2014, O’Donovan_McAuliffe_2020_exploring}. We use thematic analysis of interviews for RQ1 and RQ2, and leverage RQ1 results in RQ3---wherein we develop Slack-based PS indicators, and explore the indicators' ability to differentiate trends in PS on Slack. Our approach thus uses Creswell's \cite{Creswell_2009} sequential exploratory strategy, in which a qualitative analysis informs the development of an instrument used for a quantitative analysis. This strategy is well-suited for exploration of novel phenomena \cite{Creswell_2009}, hence its use here to inform Slack-based online PS indicators---the starting point for future automated measures.
This additionally addresses the need for PS measures that capture longitudinal \cite{Edmondson_Bransby_2023, Edmondson_Lei_2014} and behavioural data in modern, digital settings \cite{Edmondson_Bransby_2023}. Different from Ferguson and colleagues' \cite{Ferguson_Van_De_Zande_Olechowski_2024} exploration of a broad set of Slack characteristics, we pursue a targeted, mixed-methods approach, as informed by our interview findings.

\subsection{Interviews}
Since PS---a team-level construct \cite{Edmondson_1999}---needs time to emerge and stabilize \cite{Kozlowski_Chao_Grand_Braun_Kuljanin_2013, Mohammed_Ferzandi_Hamilton_2010} in design teams \cite{Miller_Marhefka_Heininger_Jablokow_Mohammed_Ritter_2019} (especially longer in virtual teams \cite{Cole_O’Connell_Marhefka_Jablokow_Mohammad_Ritter_Miller_2023}), we conduct the interviews three weeks before the semester's end. This means that teams collaborated for at least nine weeks by the interview point. The second author used Slack to recruit PMs and HRMs for 30-minute semi-structured interviews, targeting these leaders for their heightened visibility into team dynamics. Other studies also use team leaders to gauge PS at the team-level \cite{Sjöblom_Mäkiniemi_Mäkikangas_2022} and show that supervisors’ PS perceptions are significantly related to that of subordinates \cite{Frazier_Tupper_2018}.

Our institution’s research ethics board approved the interview protocol. Each participant was compensated with a \$15 USD gift card. The second author conducted the interviews in-person, using Zoom's audio-recording for transcripts. The interviews consisted of four sections. To start, participants were asked about their role(s). This was followed by questions about interpersonal risk-taking instances on their team, where the prompting examples were directly informed by Edmondson's \cite{Edmondson_1999} survey, including examples of admitting mistakes, raising concerns, mutually respecting one another, etc. Then, participants were asked about whether they perceive any difference between PS (as described) in-person and on Slack. These two settings were explicitly and repeatedly referenced in each question and follow-up to focus participants' comparisons on our study's two-setting scope. Finally, we asked about how frequently their teammates take interpersonal risks, and what team signals they perceive to affect these behaviours.

\subsubsection{Interview Sample} Among the 12 participants who consented to and scheduled an interview, there was an even split of PMs and HRMs. The sample had 1 to 5 teammates from each of the six teams. Eight participants identified as men, three as women, and one preferred not to provide a gender. Representation across races and ethnicities included White, Hispanic or Latino, and Asian. We do not break this down further due to the small sample size.

\subsection{Thematic Analysis}
We began by analyzing the data with respect to RQ1, to describe how hybrid engineering design teams perceive PS.
We pursued an inductive coding approach, different from previous studies on PS in design teams who use directed content analysis \cite{Cole_O’Connell_Marhefka_Jablokow_Mohammad_Ritter_Miller_2023, Cole_O’Connell_Gong_Jablokow_Mohammad_Ritter_Heininger_Marhefka_Miller_2022, O’Connell_Cole_Mohammed_Jablokow_Miller_2022}. This was motivated by the novelty of our RQs: an understanding of how PS manifests in hybrid teams is a gap in the literature, encouraging our data-driven 
approach in thematic analysis \cite{Braun_Clarke_2006}. Thus, we did not limit ourselves to existing PS frameworks, such as the multi-category characterization by \citet{O’Donovan_Van_Dun_McAuliffe_2020} that was then used by \citet{Ferguson_Van_De_Zande_Olechowski_2024}. During RQ1 analysis, we undertook constant comparison between our data and our operationalization of PS: a team climate characterized by \textit{interpersonal risk-taking} and \textit{mutual respect}, as informed by \citet{Edmondson_1999}. In presenting each facet of our framework in Section \ref{results:RQ1}, we will describe how the facet and its indicators are grounded in, and present an extension of, our PS definition.

The first author conducted thematic analysis in NVivo 14 software, following the multi-phase Braun and Clarke \cite{Braun_Clarke_2006} method. In \textbf{phase one}, the first author listened to all 12 audio recordings and noted any salient anecdotes, after which there was a conversation with all authors about potential themes that may arise in the coding phase. 
In \textbf{phase two}, the first author applied open codes to all transcripts, capturing perceptions of team behaviours and norms. As an example, the following snippet was associated with codes ``\textit{fear}'' and ``\textit{silence}'': ``\textit{I definitely see it with some other teammates where they don't want to speak up because they are scared of what people think...}'' This phase initially resulted in 98 codes which the first author then organized into six themes for \textbf{phase three}, then discussed with all authors in \textbf{phase four}. Five rounds of discussions occurred here, each motivating the first author to revisit the transcripts and refine the codebook by adding, eliminating, and renaming codes. The final codebook was an iterative product, resulting in five themes and 24 sub-themes for RQ1. See Appendix \ref{appendix:codebook} for the final codebook with examples. 

With an understanding of the five facets of PS in hybrid engineering design teams, we next chose to analyze perceptions of how PS differs on Slack through these facets (RQ2). We conducted a scoped coding approach (mixed inductive and deductive coding): the first author tagged all interview data about how PS differs on Slack from in-person (52 codes) as within one of the five facets, although no comments were found pertaining to differences in Leadership. Then, within each facet, the first author applied inductive coding (using the same thematic analysis approach \cite{Braun_Clarke_2006}) to identify \textit{how} that facet was perceived to differ on Slack. This comparison of two modalities necessitated a different set of sub-themes than RQ1. Discussions with the other authors resulted in two sub-themes in Communication, three in Cooperation, two in Culture, and three in Learning. The first author then reviewed all transcripts again, to identify any causes not already captured by the coding scheme, for which none were found.

\subsection{Slack Analysis}
We had access to every team’s full Slack data from all public channels, including messages, emoji reactions, replies, and metadata. We excluded all data from individuals who did not consent to sharing their Slack data with us. This brought the dataset’s total number of messages (incl. replies) from 40,833 to 38,232 (down 6\%). We present the Slack analysis details in the results section, as the interview findings directly inform the analysis' design. 

\section{Results}
We present our hybrid teams' perceptions of PS and how they differ on Slack from in-person interactions in Section \ref{results:RQ1} and Section \ref{res:how-slack-differs}, respectively. We split RQ3---our exploration of Slack-based PS indicators---into Section \ref{results:create-slack-signals} and Section \ref{res:slack-trends-heatmaps}, to first detail their design, then study their ability to unveil trends in PS on Slack.

\subsection{RQ1: The Five Facets of PS in Hybrid Engineering Design Teams}\label{results:RQ1}
\subsubsection{Preamble} We organize the interview-derived PS indicators into five facets that signal and/or encourage PS, or lack thereof. As previously discussed, PS is often described as a self-reinforcing cycle \cite{hsiang2019exploring}, where signals of PS continue to reinforce the building of PS. Thus, we refrain from categorizing these facets as either \textit{actions} to support PS or \textit{signals} of high PS, and instead use \textit{indicator} as an inclusive term. We argue that the indicators are important for leaders to monitor and encourage (or discourage, in some cases) for Psychologically Safe teams.

The five facets concern information exchanges (Communication), willingness to collaborate (Cooperation), norms and guiding principles (Culture), management behaviours (Leadership and Status), and attitudes towards and channels for feedback (Learning). 
Our five-facet organization, grounded in Edmondson's \cite{Edmondson_1999} PS operationalization, offers a new depth in illuminating underlying indicators, especially for hybrid teams. This contribution will be discussed further in Section \ref{discuss:actionable}. 
Below, we describe each of these facets and their sub-themes.


\subsubsection{Communication}
Here we present sub-themes that illustrate how mutual respect and interpersonal risk-taking manifest in the communication exchanges of hybrid teams. Interpersonal risk-taking presents in the immediacy and transparency of teammates' thoughts (\textit{delay and discreetness}), while mutual respect is observed through efforts to politely and fairly uplift everyone's voice (\textit{collective decision-making}, \textit{turn-taking}, \textit{respectful language}, and \textit{dominant voices}) (see Table \ref{table:freq-comms}).




\begin{table}[h]
\footnotesize
\caption{Communication sub-themes and the number of interviews (out of 12) that support this facet.}
\label{table:freq-comms}
\begin{tabular}{lc}
\hline
Sub-theme       & \# of Interviews\\ \hline
Delay and discreetness (negative) & 7 \\
Collective decision-making  & 6 \\
Turn-taking          & 6 \\
Dominant voices (negative) & 5 \\
Respectful language & 4
\end{tabular}
\end{table}

Starting with negative indicators, some teammates delay their thoughts, and share them only with silos with whom they are comfortable instead of the whole team. Participants attribute this discreetness to avoiding conflict or idea dismissal, unfortunately resulting in tension and inefficiencies when issues surface late and require revisiting decisions. Some teams struggle with teammates who dominate discussions, while the rest of the team is cut off from the conversation: ``\textit{We had a few strong personalities...they kind of start bouncing off each other... then the rest of the table is kind of cut off}'' [P4].


We proceed to healthy indicators, which show that team norms are nuanced and developing. We commonly hear about teams striving for greater equality in participation. While one participant discussed their team's recent turn-taking efforts, another participant from the same team perceived the efforts to be effective: ``\textit{[Our team] is getting better at making sure all voices are heard, especially in those like big 14-person meetings}'' [P5].



Some teams actively monitor their discussions to ensure respectful communication. This includes checking for teammates who want a speaking opportunity, using safe words to cut off unproductive conversations, and avoiding any implications of sole ownership when describing product parts. Participants also positively perceive teammates who confidently speak up to the whole team---doing so ensures that everyone is on the same page about key decisions being made: ``\textit{Both sides felt relatively strongly. Both sides had reasons...Everyone's like very passionate, which I think is a good thing}'' [P11].



\subsubsection{Cooperation} Here we present sub-themes that illustrate how PS (mutual respect and interpersonal risk-taking) emerge through teammates' effort in tasks, and respect for the same of their colleagues, in hybrid teams (see Table \ref{table:freq-coop}). Specifically, \textit{stubbornness} and \textit{low effort} explain that teammates must respect their colleagues' ideas and contributions, and fair division of responsibility. Mutual respect also presents in teammates' \textit{dedication}---great care and consideration for others' needs. Especially appreciated are teammates who show \textit{initiative}, despite the interpersonal risks associated with new and greater responsibility.


\begin{table}[h]
\footnotesize
\caption{Cooperation sub-themes and the number of interviews (out of 12) that support this facet.}
\label{table:freq-coop}
\begin{tabular}{lc}
\hline
Sub-theme       & \# of Interviews \\ \hline
Initiative         & 5 \\
Low effort (negative) & 5 \\
Dedication & 4 \\
Stubborn (negative) & 3  
\end{tabular}
\end{table}

Starting with the negative indicators, inadequate effort sometimes takes the form of weaponized incompetence---a teammate will initially take responsibility for a task, but attempt to offload it later, claiming not to be skilled enough for the task. This issue may be a product of time pressures: ``\textit{As the semester has gone on, it has been a little bit of ‘Hey, can you do this?’ It's like, `I'm happy to work with you to learn how to do it', and then they're like, `Actually, maybe not}'' [P12].
Teams may also struggle with teammates who disregard how their actions impact the team. Tensions start to rise when teammates are intensely attached to their ideas or contributions. There may also be inattentive teammates whose careless mistakes lead to duplicated efforts: ``\textit{There's like no limit to `Oh, I'll just go in and mess with that}'... '' [P10].    


Healthy Cooperation indicators include teammates who are willing to help each other out. This includes covering for each other when outside commitments or mental health issues get in the way: ``\textit{We've made it very clear that, if you're ever out of bandwidth, for whatever reason.... mental health, extra-curriculars...We're like, `that's totally fine... We will find somebody else to cover your work'...} '' [P5].
Next, initiative is described as being shown by teammates who actively work towards being more useful: ``\textit{The [teammates] who had no experience in electrical, they were really helpful in developing the battery box... Then after that, they were like, `Okay, now, I can be more useful elsewhere' and found somewhere else [to be] pivotal}'' [P11].



\subsubsection{Culture} Within Culture, sub-themes that support interpersonal risk-taking include team-bonding efforts that strengthen the relationship between team members \textit{(community)} such that they feel safe to take risks, and gracious responses to mistakes to mitigate fear of negative consequences (\textit{no blame}). An aversion to interpersonal risk-taking may manifest in calculated behaviour (\textit{fear}), while mutual respect may emerge through consideration for teammates' efforts and capabilities (\textit{gratitude}) (see Table \ref{table:freq-culture}).

\begin{table}[h]
\footnotesize
\caption{Culture sub-themes and the number of interviews (out of 12) that support this facet.}
\label{table:freq-culture}
\begin{tabular}{lc}
\hline
Sub-theme       & \# of Interviews        \\ \hline
Community   & 8 \\
No blame          & 8\\
Fear (negative) & 6\\
Gratitude & 5
\end{tabular}
\end{table}

Regarding fear, teammates may be averse to failure and concerned with impression management: ``\textit{I definitely see it with some teammates where they don't want to speak up because they are scared of what people think...they want to seem chill...[not] uncool}'' [P4].

Moving onto positive sub-themes, participants describe being committed to collective accountability over blame, since blame can lead individuals to withdraw or disengage: 

\begin{quote}
Shame [has] the tendency of really shutting people off. Someone may say, `Oh, you only think of me as some leach'... `No, you're actually a very smart person. We just want more out of you, because we know there's more for you to give'. [P8]
\end{quote}

Next, we acknowledge an emphasis on gratitude. While usually explicit (e.g., saying ``\textit{thank you}''), gratitude may also be implicitly achieved. For example, a fair division of labour contributes to one participant’s sense of being a valued contributor: ``\textit{I think we're assigning each other tasks that feel valuable…not busy work. It's like they're trusting me with things to do that are important}'' [P9].  

Additional activities that may strengthen, or indicate strong, interpersonal relations include team-bonding activities that foster safe spaces and remind teammates that they can be friends: ``\textit{We tried to do team events [and] activities...We definitely established a culture that was a safe space where people felt comfortable sharing}'' [P11].

\subsubsection{Leadership and Status} Participants additionally relate their perceptions of PS (interpersonal risk-taking and mutual respect) to team leadership and hierarchy (see Table \ref{table:freq-leadership}). \textit{Disengaged leaders} here are a negative PS indicator, potentially threatening PS by failing to uphold, or monitor for, interpersonal risk-taking and mutual respect. Respect may also lean in favour of certain teammates, instead of being equally distributed (\textit{social hierarchy}). A team may exhibit \textit{overwhelmed leaders} when non-leaders defer or avoid nontrivial responsibilities, placing an unfair and potentially disrespectful burden on PMs. \textit{Shared leadership} demands that the interpersonal risks associated with being a leader be distributed, allowing individuals to influence the team and foster impact---regardless of their formal title.



\begin{table}[h]
\footnotesize
\caption{Leadership sub-themes and the number of interviews (out of 12) that support this facet.}
\label{table:freq-leadership}
\begin{tabular}{lc}
\hline
Sub-theme       & \# of Interviews\\ \hline
Overwhelmed leaders (negative)  & 6\\
Shared leadership  & 4 \\
Social hierarchy (negative) & 2\\
Disengaged leaders (negative) & 1
\end{tabular}
\end{table}


In negative indicators, we observe weak leadership, overwhelmed leaders, and social hierarchies. For example, one HRM perceives an unwritten social hierarchy across subteams. Due to their elevated status, the HRM feels the need to be cautious and hesitant in their interactions with two sub-teams: ``\textit{For some reason, there is a hierarchy with [the] design and mechanical teams, where [the broader team] think[s] [these two teams] are more important, which is annoying, and so I need [these two teams] to defend me}'' [P10].

PMs may also be perceived as carrying a disproportionately heavy share of the workload: ``\textit{We had a big thing where the [PMs] got overwhelmed and [other teammates] came out saying `I didn't have that much work to do. I could have done that' but they didn't say [anything earlier]}'' [P4]. Contrasting overwhelmed leaders, we also learn about leaders who may be wholly disengaged, albeit less frequently.
Otherwise, leadership behaviours that are positively perceived concerning PS encourage honest participation and equal contributions. In teams with such norms, everyone is empowered to influence and exercise autonomy: ``T\textit{here's a third [sub-team] that doesn't have any PMs, and work was much more evenly distributed... It did end up helping overall}'' [P1].


\subsubsection{Learning}This facet shows our hybrid teams' innovative context (\textit{engineering design}) linking PS, as operationalized, to learning-related indicators (see Table \ref{table:freq-learn}). Sub-themes closely aligned with interpersonal risk-taking include \textit{seeking help} and embracing \textit{new ideas}, \textit{critiques}, \textit{mistakes}, and \textit{obvious questions}. \textit{Teach} encompasses efforts to have teammates learn, instead of offload, difficult tasks, again indicative of risk-taking. Respect is perceived through a team’s willingness to take seriously calls for significant change and improvement in team processes (\textit{adapt and improve}).



\begin{table}[h]
\footnotesize
\caption{Learning sub-themes and the number of interviews (out of 12) that support this facet.}
\label{table:freq-learn}
\begin{tabular}{lc}
\hline
Sub-theme       & \# of Interviews\\ \hline
Embrace critique  & 11\\
Adapt and improve       & 9\\
Embrace mistakes & 8\\
Seek help & 7\\
Teach & 7 \\
Embrace new ideas & 6\\
Embrace obvious questions & 5
\end{tabular}
\end{table}

We primarily hear about teams demonstrating these sub-themes positively. Nonetheless, some participants perceive their teammates to withhold critiques, in which case anonymous feedback mechanisms may be the solution for unveiling honest opinions: ``\textit{I forced everybody to sit down for five minutes and write out the problems that they saw… anonymously…Everyone said… `I did not appreciate having to do all the work Sunday night...' [I] was like...we're getting somewhere}...'' [P10]. Otherwise, we hear about teammates openly critiquing ideas during team-wide meetings. Participants credit such norms to presenting ideas without implying full confidence in their correctness: ``\textit{I try not to like be like, `Oh, this is my method [and] my method is correct.' I think a lot of people on the team also feel this way...`These are my ideas [rather]}'...'' [P6]. 

Teammates also admit their limitations and ask for help, which is positively received. Teammates who seek help are met with another teammate taking over the task when the team is short on time, otherwise teammates are directed to learn from someone else. 
Intra-team teaching efforts may also be planned and intentional---teams may curate sub-teams with a mix of expertise levels. 
\begin{quote}
We had a lot of people want to grow in electronics and CAD, so [we tried] to get people without CAD experience with people [that have it]...It's been working as well as it can. It's like a little tough with stuff moving so quickly. [P5]
\end{quote}
Next, we hear about teams refraining from directing any ill-will towards those who ask obvious questions, or putting too much stock into mistakes. Participants go so far as to describe the acceptance of mistakes as only natural and necessary during fast-paced development: ``\textit{By the nature of the project...People are just making mistakes, left and right. We're all learning... It's really fast and rapid. It's not `Oh, no, I made a mistake}'...'' [P7].

We additionally hear about improvement orientations in teams. Teams are proactive with feedback and forgo familiar processes in favour of evolving needs: ``\textit{We rearranged task forces... We also had a four-hour-long meeting to discuss our idea-choosing process, 
and it was super helpful [for producing an] inclusive idea-picking process}'' [P1].

\subsection{RQ2: Perceptions of How PS Differs on Slack from In-person}
\label{res:how-slack-differs}
Here we present how hybrid teams perceive PS to differ on Slack from in-person collaboration in relation to Communication, Cooperation, Culture, and Learning (four of the five facets). Where expressed, we also report participants' perceptions of causal mechanisms behind the differences.

\subsubsection{Communication is reserved and ambiguous}
While some people speak more on Slack, 
reports of teammates speaking less are most common. Participants reported that teammates are cautious in what they say on the platform since a Slack message leaves a paper trail (i.e., saved and searchable record), and feels more effortful to respond to than in-person speech: ``\textit{It just feels like, `Is it necessary for me to ask?' because [to me] and others... it feels like a Slack message has a larger weight than speaking...}'' [P6].

Slack communication is also sometimes described as unclear and inefficient. According to P7, Slack messages are subject to long response times and high cognitive load from considering correct grammar use and deciding which channel to use. Participants also reference challenges in expressing tone: ``Y\textit{ou can't hear someone's vocal tones. You can't see the difference between `Oh, I'm doing fine' or `I'm doing fine-ish'...}''[P12].

\subsubsection{Cooperation is deficient and doubted}
Teammates perceive weaker commitment from their colleagues on Slack who are inactive and unwilling to read long Slack threads: ``\textit{People also
don't react well to there being a thread of like 50 messages. They're like, `I'm not reading all that.' It's like, `come on'}... '' [P5].
Additional issues include teammates who are less willing to help, trust their colleagues, and compromise over Slack. For example, teammates may be more dismissive of requests for support: ``\textit{I guess the tone is a bit more, `I would just check the CAD' on Slack. I think people are more receptive to being a bit more helpful in-person}'' [P2].

Teammates are also less trusting of each other's intentions on Slack. This extends to requests for help, which are met with doubt about whether a ‘good faith’ effort was first attempted:
\begin{quote} If I was in-person, then I could show them like `Hey, I've tried this and this' super easily...It's a lot harder to do over Slack. It's hard to tell, `Okay, like, did you actually try? Or is this just like a cop-out?' [P9]
\end{quote}

\subsubsection{Culture is conservative}
Teams perceive Slack to be mostly task-oriented since it is challenging to read each other's emotions on the platform. P11 describes that their team relies upon in-person meetings to anchor rich discussions and uses Slack for more straightforward purposes later (e.g., voting).
According to P8, teammates can be inauthentic on Slack, especially in the \#general channel: ``\textit{In-person, the willingness to say something stupid is a lot higher...whereas, if you're in the \#general [channel on] Slack…it feels more of a performance than it feels like communicating...}'' [P8].

\subsubsection{Learning is difficult and delayed}
Teammates perceive learning as less frequent and effective on Slack because of the additional effort required. We hear about members delaying critiques to in-person meet-ups---frustrating for teammates whose feedback requests are met with low engagement until they are brought up in person. We also learn that it is more difficult to critique drawings on Slack and ask questions, resulting in mistakes being caught late. 

\begin{quote}
They got it over Slack and said, `We'll go with this.'... I was like, `Does this look like a state diagram? No, it looks like a list of features'...It seems that things are kind of forced to come up in-person, whereas on Slack, people can just stand back. [P10]
\end{quote}

Similarly, it is harder to work through mistakes on Slack, sometimes because teammates differ in how much thought they put into their Slack messages: ``\textit{People hit send at different levels of thought development. I'll draft my messages with a lot of care... I know other people just like hit enter...}'' [P3].

Critiques are also withheld to avoid being misinterpreted, proven wrong, or accused of hypocrisy: ``\textit{I didn't want to send any messages as a response because I didn't know if I had done the same [thing]}'' [P1].
Similarly, participants find it is easier to generate ideas in person because they can pick up on the body language and/or tone of their colleagues, and accordingly elaborate further or pivot. These limitations of Slack communication may encourage unnatural and minimal critiquing activities. For example, teammates may request feedback on a draft critique before pressing send: ``\textit{He actually ran the message by me… He wanted to make sure he wasn't being like too accusatory… [and] that it was coming across the right way}'' [P9].

Participants also describe design discussions as being more frustrating over Slack and more effective when they happen in-person. As a result, teams move discussions to in-person settings: 

\begin{quote}
When you're having a discussion or debate and multiple people are chiming in as you're typing, you're not reading what has come up since. It's a little messy... Slack's not the ideal location for some of these design discussions... Things get missed. [P11]
\end{quote}


\subsection{RQ3a: Design of Slack-based PS Indicators}\label{results:create-slack-signals}
We found that teams perceive differences in the extent to which indicators across four PS facets are effectively expressed on Slack compared to in-person. As such, we are motivated to design PS indicators that set the stage for future automated PS monitoring, enabling teams to track and reflect on their online PS. For each of the five PS facets, we select the three most popular sub-themes, as measured by the number of interviews in which they were present, resulting in 15 Slack indicators (see Table \ref{table:res-interviews-to-slack}).
\begin{table}[h]
\footnotesize
\caption{The 15 Slack-based indicators and their sub-components. Starred (*) components are reverse-scored since they contribute negatively (positively) to positive (negative$\dagger$) PS indicators. Two indicators ($\ddagger$) are in the range $[0, 1]$. Other indicators are computed in standard deviation units, and thus do not have fixed ranges. Tools represent the packages used to access the source dictionaries and methods.}
\label{table:res-interviews-to-slack}
\begin{tabular}{p{0.13\linewidth}|p{0.23\linewidth}|p{0.31\linewidth}|p{0.10\linewidth}|p{0.08\linewidth}}
\hline
Facet & Sub-theme  & Slack Indicator Component(s) & Source(s) & Tool(s) \\ 
\hline
Communication 

& Delay and discreetness$\dagger$ & *Time until first reply, number of Slack threads, number of replies & \cite{riedl2017teams, muller2023slack, anderson2019winning, Wang_Wang_Yu_Ashktorab_Tan_2022} & N/A
\\
& Collective decision-making & `Negotiation/coordination' language  & \cite{stewart2019say} & \cite{boyd2022development} 
\\
& Turn-taking$\ddagger$& $1-$ Gini coefficient for number of messages and words  & \cite{tausczik2013improving} & \cite{hu2024-toolkit}
\\ 
\hline

Cooperation           
& Initiative & Verbs, *word count, *time-related words & \cite{make_it_happen_2021, buseyne2024peering} & \cite{boyd2022development}
\\
 & Low effort$\dagger$ & *Personal pronouns, assent, definite articles & \cite{nguyen2016effects, nguyen2014lexical} & \cite{boyd2022development} 
\\
& Dedication  & Positive emotion, *inhibition words & \cite{rand2015collective} & \cite{boyd2022development, pennebaker2007}
\\ 
\hline
Culture               
& Community  & Mimicry, word count, *first-person plural pronouns & \cite{gonzales2010language} & \cite{hu2024-toolkit}
\\
& No blame   & Positive emotion, causation words & \cite{doherty2023-humanness, stewart2019say} & \cite{boyd2022development} \\
& Fear$\dagger$ & Word count, first-person plural pronouns, second-person pronouns, certainty words, *first-person singular pronouns  & \cite{van2021language, van2019language} & \cite{boyd2022development}
\\ \hline
Leadership and status              
& Overwhelmed leaders$\dagger$ & Difference between non-PMs' \& PMs' \textit{low effort$\dagger$} scores & \cite{nguyen2016effects, nguyen2014lexical} & \cite{boyd2022development}\\
& Shared leadership   & *Standard deviation in *auxiliary verbs, numbers, affiliation words, *quotation marks & \cite{korner2024language} & \cite{boyd2022development} \\
& Social hierarchy$\dagger$ & *Standard deviation in word count, emotional tone, *anger words, *tentative words, *filler words & \cite{korner2024language} & \cite{boyd2022development} \\ \hline
Learning              
& Embrace critiques   & Word count, exclusive words & \cite{becker2022leadership} & \cite{boyd2022development} \\
& Adapt and improve $\ddagger$  & `Forward flow' & \cite{gray2019forward} & \cite{hu2024-toolkit} \\
& Embrace mistakes    & *Words longer than six letters, present-tense verbs, discrepancy words, first-person singular pronouns, *articles & \cite{shahane2019predicting} & \cite{boyd2022development, pennebaker2007} 
\\ 
\hline
\end{tabular}
\end{table}

\subsubsection{Preamble}We find links between linguistic components and measures in the literature and our interview sub-themes to inform the design of 15 indicators. 14 of them (all but \textit{adapt and improve}) are composite measures whose sub-components we convert to standardized z-scores before averaging together, as done by other discourse analysis research (e.g., \cite{shahane2019predicting, ferguson2023we, graesser2014coh}). We standardize to avoid the nonsensical means that would result if sub-components of dissimilar distributions (e.g., quotation marks vs. verbs) were averaged together in a composite measure.

As done by \citet{shahane2019predicting}, we reverse-score sub-components that are negatively related to a given indicator (i.e., multiply by $-1$) before they are averaged with other sub-components. Most of our indicators require only team-level calculations, except for \textit{turn-taking} (in Communication) and all three Leadership indicators, which involve initial teammate-level computations. Each indicator is calculated daily, and the daily values use all messages sent within the past seven days, to ensure that we have enough messages for a robust score computation: the average number of messages sent per teammate (team) every seven days is 33 (381)---which we consider sufficient sample size to assume normality and apply z-score standardization \cite{kwak2017central}. Below we describe the indicators, that are then applied in Section \ref{res:slack-trends-heatmaps}.

\subsubsection{Communication} 
\textit{Delay and discreetness} represents teammates' reluctance to freely speak up in the presence of all teammates, instead withholding or postponing their thoughts. Here we design a composite score comprising each team's: 1) average time until the first reply, 2) total number of Slack threads, and 3) total number of replies. As \citet{riedl2017teams} describes, strong temporal coordination can indicate ``higher responsiveness of activity among members of the team,'' \cite[p. 19]{riedl2017teams}, hence our use of reply times to capture the extent of \textit{delay}. Regarding the latter two components, researchers consider Slack threads as markers for team discussion \cite{Wang_Wang_Yu_Ashktorab_Tan_2022, muller2023slack}, but also use replies to quantify active engagement in online conversation \cite{anderson2019winning, muller2023slack}. Thus, we use both components to signify outspokenness in a team-wide setting, especially since our data comprises messages from public Slack channels---the antithesis to \textit{discreet} communication. All components are z-scored, reverse-scored (where needed) and then averaged to produce each team's score.

From \textit{collective decision-making}, we heard participants praise passionate discussions that ensure everyone's say in decisions being made. Here we rely on a CSCW study by \citet{stewart2019say}, in which `negotiation/coordination' is defined as a sub-activity of collaborative problem-solving, where teammates iteratively discuss, reach, and revise solutions \cite{stewart2019say}. Both `negotiation/coordination' language and our sub-theme focus on action-oriented and productive discourse for team-wide matters; thus, we use linguistic components strongly correlated with the former \cite{stewart2019say}. For each team, we compute messages' z-scored frequency of words from each of the `negotiation/coordination' components\footnote{the Linguistic Inquiry and Word Count (LIWC) categories \cite{boyd2022development} that we use from \citet{stewart2019say} are causation, negations, insights, auxiliary verbs, articles, differentiation, and personal pronouns.}, resulting in seven values averaged to produce one team score.

\textit{Turn-taking} represents goals for equality in conversational participation. We use the equal participation feature from \citet{hu2024-toolkit}, originally proposed by \citet{tausczik2013improving}. The feature computes the Gini coefficient (0--1), where (0) 1 indicates perfect (equality) inequality in the number of contributions (e.g., messages) across teammates. We compute for each team their \textit{reversed} equal participation score for number of words and number of messages, averaging the two scores. In weighing words and messages equally, the score does not favour many short messages over a few lengthier messages, or the reverse.

Of the Communication indicators, only \textit{turn-taking} incorporates teammate-level analysis (with the Gini coefficient---a measure of dispersion), but a future measure could incorporate teammate-level analysis in the first two indicators, too. At the same time, teammate-level analysis may unnecessarily complicate the first two indicators. With the `time until first reply' component in \textit{delay and discreetness}, any between-teammate differences in how fast messages receive their first reply go beyond the spirit of the Communication indicators, and start to leak into the spirit of the Leadership indicators (e.g., \textit{social hierarchy})---which \textit{will} make use of teammate-level analyses. As for \textit{collective decision-making}, we exclude teammate-level analysis under the assumption that the presence of its linguistic components in public Slack channels (in teams of 14-16 members) sufficiently signals team-wide involvement and engagement in the conversation. If a few teammates are dominating the conversations here, however, our \textit{turn-taking} metric should capture such dynamics. We ultimately restrict teammate-level analyses to four metrics whose definition mandates it: \textit{turn-taking} and the three Leadership indicators (still to be introduced).
 \subsubsection{Cooperation} \textit{Initiative} refers to teammates who actively seek new and improved ways to contribute. We use linguistic components associated with showing agency \cite{make_it_happen_2021} and being execution-driven, regularly taking on and reliably completing tasks \cite{buseyne2024peering}. Indicators include verb use \cite{make_it_happen_2021}, but less time-related words and word count overall, likely since proactive teammates favour immediate action over prolonged discourse \cite{buseyne2024peering}. For each team, we compute messages' z-scored frequency of total word count, verbs, and time-related words, reverse-scored (where needed) and averaged.

\textit{Low effort} represents teammates' tendency to offload, neglect, and/or trifle their responsibilities. To measure this sub-theme, we rely on prior conversational analyses of uninvolved and distracted teammates, who are more likely to use personal pronouns, and less likely to use assent words and definite articles, suggestive of their inward focus, inattention, and cognitive detachment to ideas and objects being discussed, respectively \cite{nguyen2016effects, nguyen2014lexical}. Thus, we computed the z-scored frequency of each team's messages' personal pronouns, assent words, and definite article words, reverse-scored (where needed) and averaged.


\textit{Dedication} captures the extent to which teammates cater their collaborative efforts to honour the workload and needs of their colleagues. \citet{rand2015collective} found that cooperative and selfless behaviours during games were significantly predicted by high positive emotion and low inhibition (e.g., ``constrain'', ``stop'') language. As a result, we compute messages' z-scored frequency of positive emotion and inhibition words for each team, reverse-scored (where needed) and averaged.
\subsubsection{Culture} \textit{Community} entails efforts to foster inclusion and strong intra-team bonds. We rely on linguistic components identified as indicators of group cohesiveness---positive intra-team dynamics, social synchrony, and strong rapport \cite{gonzales2010language}. Mimicry of function words (auxiliary verbs, articles, etc.) and word count are significant, positive predictors of cohesion, whereas first-person plural pronouns use is a significant, negative predictor, potentially because teammates use such pronouns to linguistically compensate for a lack of community \cite{gonzales2010language}. We use messages' z-scored frequency of words, mimicry score from \citet{hu2024-toolkit}---which uses cosine similarity between message embeddings to quantify the invariance in team language style---and first-person plural pronouns, reverse-scored (where needed) and averaged.

\textit{No blame} is a commitment to accountability without blame and shame. Detecting blame and other negative micro-behaviours in text is a challenging task, sometimes producing non-intuitive results \cite{paromita2023linguistic}. As such, we focus on healthy accountability indicators: positive tone \cite{doherty2023-humanness} and causation word use \cite{doherty2023-humanness, stewart2019say} correspond with healthy team processes and reflective, reasoning-related thinking, respectively. We thus compute each team's z-scored frequency of these two categories, then average the two values.

\textit{Fear} describes teammates who are overly concerned with impression management and risk of failure---being highly cautious and calculated in what they say or do. Contrasting \textit{fear} would be teammates who exhibit confidence in their skills and ideas. Teammates who rate themselves high in confidence and credibility, and openly express opinions outside the team's status quo, use language with more words, first-person plural and second-person pronouns, and certainty, but fewer first-person singular pronouns \cite{van2021language, van2019language}. Hence, we measure each team's messages' z-scored frequency of these sub-components, reverse-scored (where needed), and averaged.
\subsubsection{Leadership and Status}
\textit{Overwhelmed leaders} refers to perceptions that the PMs are investing more time and effort on tasks than their teammates, to an extent that others deem unfair. We rely on our previous \textit{low effort} indicator to capture this contribution disparity. Recall that \textit{low effort} considered (personal pronouns) assent words and definite articles as (negative) indicators of involvement and focus, as informed by research on uninvolved and distracted teammates \cite{nguyen2014lexical, nguyen2016effects}. Other work has also proposed excluding first-person plural pronouns to study task-focused (as opposed to self-focused) discourse in teams \cite{hu2024-toolkit, tausczik2013improving}. Using these insights, we compute for two groups---PMs (two teammates) and then non-PMs (i.e., the rest of the team)---their \textit{low effort} score, then compute their difference such that negative scores indicate more effort in PMs.

\textit{Shared leadership} encompasses a norm where leadership and ownership are distributed among all team members rather than resting solely with the official leaders. The nature of this indicator aligns with the literature on shared leadership, in which decentralized power and influence are described as defining characteristics \cite{zhu2018shared}. As a result, we use linguistic components that significantly predict perceived workplace power and social influence to compute `power' scores for each teammate. We compute for each teammate their messages' z-scored frequency of auxiliary verbs, numbers, affiliation words (e.g., ``we'', ``help''), and quotation marks \cite{korner2024language}. After reverse-scoring (where needed), we average components.
To measure dispersion among these scores, we compute the standard deviation of the `power' score distribution, as opposed to relying on the Gini coefficient, since the sum of `power' scores is not meaningful. Since lower scores suggest less dispersion, the team's standard deviation is reverse-scored.

\textit{Social hierarchy} encompasses perceptions that teammates of certain affiliation are elevated in status. This aligns with the Prestige scale by \citet{cheng2010pride}, which measures the extent to which an individual has been granted status by others, as a result of their perceived superiority in expertise and potential for success. Using this scale, \citet{korner2024language} identifies linguistic predictors of prestige. Based on \citet{korner2024language}, we compute `prestige' scores using an individual's messages' z-scored frequency of word count, emotional tone, anger words, tentative words, and filler words, reverse-scored (where needed) and averaged. Similar to \textit{shared leadership}, we compute the standard deviation among prestige scores. As \textit{social hierarchy} is a negative PS indicator, we reverse-score the resulting standard deviation, for higher scores to indicate less disparity in prestige.
\subsubsection{Learning}
\textit{Embrace critiques} entails openly challenging each other's ideas. We use the linguistic components that \citet{becker2022leadership} uses to measure `intellectual stimulation' in messages, which ``involves challenging the status quo, questioning [assumptions], and suggesting alternative perspectives'' \cite[p. 382]{becker2022leadership}, as defined by Bass \cite{bass1985leadership}. This means we compute messages' z-scored frequency of exclusive words (e.g., ``except'') and total word count, then average the scores.

\textit{Adapt and improve} encompasses an improvement orientation, where teammates adapt processes and policies to meet new, evolving needs. Here we rely on the `forward flow' score from \citet{hu2024-toolkit}, originated in \citet{gray2019forward}. To quantify the extent to which ideas in text are evolving (as opposed to static and similar), the `forward flow' score at a particular message uses the cosine distance between the message's vector embedding and the averaged vector embedding of all preceding messages \cite{hu2024-toolkit}. Values closer to 0 (1) indicate minimal (high) topic evolution \cite{hu2024-toolkit}.

\textit{Embrace mistakes} entails perceptions that mistakes are frequent, inevitable, and invaluable amid fast-paced innovation. 
We use linguistic components related to `psychological closeness' (low `psychological distance')---described as use of charismatic and honest (as opposed to distant and deceptive) language in addressing the team's reality \cite{becker2022leadership, p-crowdfund}. Since individuals harness psychologically distant language to withdraw from present circumstances \cite{p-crowdfund}, psychological closeness is more common in teams who authentically admit and acknowledge mistakes and associated consequences. Adapting Shahane and Denny's \cite{shahane2019predicting} psychological distance measure, we compute each team's z-scored frequency of present-tense verbs, discrepancy words (e.g., ``could'', ``should''), first-person singular pronouns, words longer than six letters, and articles, reverse-scored as in Table \ref{table:res-interviews-to-slack}, before averaging. Note that we reverse-score the components related to articles and long words since they signal impersonal and detached, not personal and present-focused, language \cite{shahane2019predicting}.

\subsection{RQ3b: Exploratory PS Trends on Slack as Suggested by the Slack-based PS Indicators}
\label{res:slack-trends-heatmaps}
\begin{figure}
\centering
\includegraphics[scale=0.345]{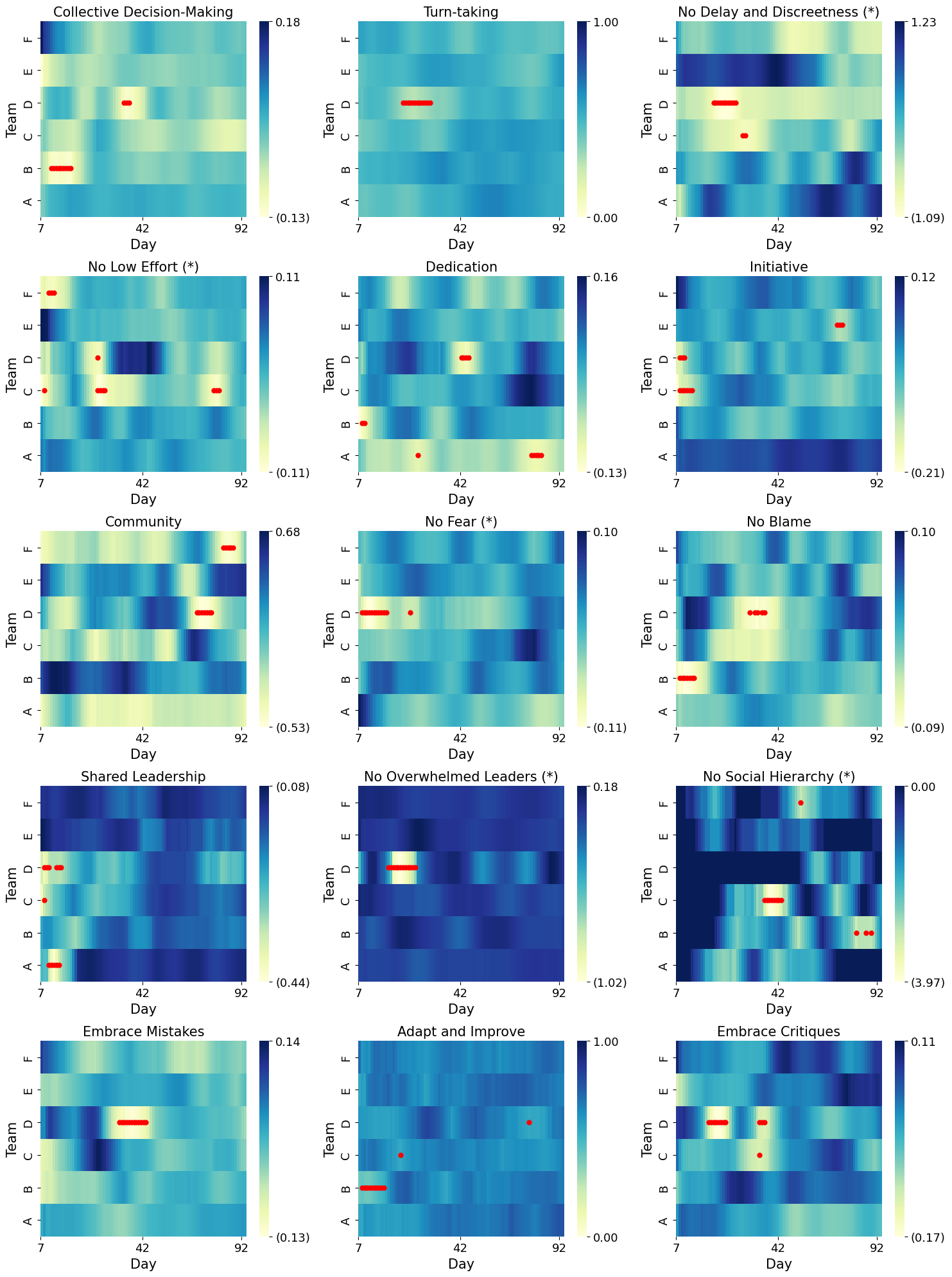}
\Description{X}
\caption{These heatmaps present teams' relative performance in each indicator across the five facets: Communication, Cooperation, Culture, Leadership, and Learning from top to bottom. Time (x-axis) starts at Day 7 since every computation uses messages sent in the past week. Darker (lighter) colours suggest more (less) Psychologically Safe performance. The red circles are data points in the bottom 2.5\% for the particular indicator. Starred (*) titles mark originally negative PS indicators, reversed here for consistency in the plots. Most ranges are nearly symmetrical about zero except for \textit{turn-taking}, the three Leadership indicators, and \textit{adapt and improve}---five indicators whose calculations differ slightly from the rest (see Table \ref{table:res-interviews-to-slack}).}
\label{fig:heatmaps}
\end{figure}

Figure \ref{fig:heatmaps} shows teams' \textit{relative} performance by indicator, over time and compared to each other. We avoid implications of objectively `good' or `bad' scores. Instead, we focus on \textit{relative} performance, from which we find that our indicators capture variance between and within teams across facets and indicators.
Since teams perceive differences in the extent to which indicators across four PS facets are effectively expressed on Slack compared to in-person, as per Section \ref{res:how-slack-differs}, the trends do not reflect teams' ``complete'' PS. Rather, they reflect teams' \textit{relative} PS \textit{on Slack}. To be discussed in Section \ref{sec:discussion-slack-signals}, these indicators satisfy the need for PS measures that are longitudinal \cite{Edmondson_Lei_2014, Edmondson_Bransby_2023} and catered to nontraditional environments \cite{Edmondson_Bransby_2023}. They ultimately set the stage for future automated PS measurement on Slack and related platforms, but must be first followed by in-depth empirical work for validation and user-centred design changes. Example modifications that result might include more advanced data analysis, such as \textit{no blame} measurement using only messages that discuss mistakes, or \textit{low effort} measurement using only task-related communication. However, such modifications can introduce unintended biases, underscoring the need for the Slack-based PS indicators' robust validation (to be discussed in Section \ref{sec:limits}). Herein, our application of these indicators serves to explore their ability to elucidate PS trends and differentiate teams---not to make definitive claims about our six teams' PS on Slack.

\subsubsection{Observations} We first compare trends \textit{within} each facet, which suggests that indicators are complementary in their contributions to team insights. We first notice this in Communication, in which the patterns for \textit{delay and discreetness} and \textit{turn-taking} are similar (see Team C and Team D), whereas trends in \textit{collective decision-making} differ, including Team B's initial, extreme dip. We speculate this to be a result of \textit{collective decision-making}'s emphasis on participation \textit{content}, contrasting the other two indicators' strict focus on speed and volume metrics. Cooperation serves as another example, in which Team A has the best (worse) \textit{initiative} (\textit{dedication}) scores. Recall that \textit{initiative} encompasses teammates' intentional pursuit for more responsibility (Section \ref{results:RQ1}). Potentially, there is a \textit{dedication--initiative} trade-off, where those who exhibit high \textit{initiative} expect their teammates to do the same and thus appear apathetic to their team's needs over Slack (low \textit{dedication}). In Culture, \textit{community} sees the most inter-team variance. This might suggest that strong interpersonal bonds (\textit{community}) are more effortful than the absence of \textit{fear} and \textit{blame}, thus higher in discriminative power across teams. Regarding Leadership, we notice that \textit{shared leadership} and \textit{no overwhelmed leaders} are not identical, reinforcing the distinction between concentrated power/influence and excessive task delegation to official leaders. Learning shows teams mostly indistinguishable in \textit{adapt and improve}, in contrast to the other two indicators. Interim in-person discussions (not captured in Slack data) might explain why Slack conversations evolve at a consistently high rate, suggesting another measure might be more useful here.

There are also notable \textit{between}-facet trends, suggesting that these facets capture different elements of a complex construct. First, the lowest scores are always observed \textit{before} the midpoint in Communication, whereas their distribution is scattered in other PS facets. This might indicate that teams overcome their worst struggles in Communication more quickly than the other, potentially more involved, facets. We also notice that the most extreme fluctuations in performance are reserved for Leadership, which comprises three indicators that all incorporate teammate-level analysis (not merely team-level). Most dramatic is \textit{overwhelmed leaders}, in which Team D sees the only steep dip. As this indicator measures the difference between PM and non-PM \textit{low effort} scores, the infrequent negative values mean PMs rarely exert more effort, potentially because they play a facilitative role on Slack. Ultimately, the more dramatic trends in Leadership might result from role-specific nuances that are less relevant to other facets. Furthermore, differences across facets reinforce the distinction between similar indicators. The trends in \textit{embrace mistakes} (under Learning) differ slightly from that of \textit{no blame} (under Culture): Team D relatively struggles in both, whereas Team C's relative performance is stronger in the former. This suggests that these two indicators are related but distinct, as we would expect: responding graciously to errors (\textit{no blame}) is different from actively harnessing errors as learning opportunities (\textit{embrace mistakes}). We end here by drawing attention to the team that hosts the least Psychologically Safe scores in almost all indicators: Team D. This might suggest their comparatively greater instability.




\subsubsection{Takeaway}
This analysis is intended to illustrate the potential use of these Slack indicators for future monitoring of online PS on Slack. As shown, their deployment differentiated the team with the most extreme dips in performance (Team D, in this case), and identified trends between and within facets. We provide some evidence that these 15 indicators complement each other to inform how PS presents on Slack over time. Our indicators may be a start for future monitoring systems that not only track dips in PS, but also identify which team(s) are at risk, when, and for which indicator(s), to then suggest how the indicator(s), and PS on Slack more broadly, can be improved.

\section{Discussion}
In exploring PS in hybrid engineering design teams, we identified a five-facet framework with 24 indicators grounded in Edmondson's \cite{Edmondson_1999} operationalization (RQ1); four perceived differences in PS on Slack compared to in-person (RQ2); and 15 Slack-based PS indicators (grounded in RQ1 findings) that we show differentiate teams across PS indicators and facets (RQ3). We proceed to discuss our three contributions to HCI and PS literature.

\subsection{Actionable PS Framework for (Hybrid) Teams Grounded in PS Theory}
\label{discuss:actionable}
With RQ1, our investigation of hybrid engineering design teams' perceptions of PS indicators resulted in a detailed, five-facet PS framework comprised of 24 total PS indicators. Grounded in Edmondson's \cite{Edmondson_1999} PS operationalization---all 24 indicators were perceived as being associated with the concepts of \textit{interpersonal risk-taking and mutual respect}, but distinguish themselves as being more granular and actionable than high-level and conceptual. We fill a gap, where previous PS research falls short in identifying specific words and actions linked to PS \cite{Edmondson_Bransby_2023}. We thus enable hybrid teams---otherwise overlooked in PS research---to translate an abstract concept to behaviours worth encouraging along the universal dimensions of Communication, Cooperation, Culture, Leadership, and Learning.

\subsubsection{Hybrid Teams}
We contribute to emerging HCI research on hybrid teams a framework of behaviours whose implementation and prioritization may buffer, if not remedy, some of these teams' known challenges. Our framework complements Liu and colleagues' \cite{Liu_Van_Essen_Eggen_2024} design space for improved social awareness and informal communication in hybrid teams, in that our PS indicators elaborate on the types of \textit{activities} (e.g., language) that may function as socially meaningful information cues. To the findings in \citet{Bjorn_Busboom_Duckert_Bødker_Shklovski_Hoggan_Dunn_Mu_Barkhuus_Boulus-Rødje_2024} that show hybrid teams enduring difficulties in relationship- and trust-building, we contribute a framework of indicators that are grounded in a construct (PS) that encompasses team climates of interpersonal trust and mutual respect \cite{Edmondson_1999}. Going beyond support for hybrid teams' relational work, some of our indicators, such as \textit{social hierarchy} (under Leadership and Status) and \textit{dominant voices} (under Communication) serve as additional explanations for, or manifestations of, the relational asymmetries in hybrid teams. Thus, these teams' asymmetries may not depend solely on their multi-location nature \cite{Bjorn_Busboom_Duckert_Bødker_Shklovski_Hoggan_Dunn_Mu_Barkhuus_Boulus-Rødje_2024}, and instead depend, in part, on PS.

\subsubsection{PS Literature}
To PS literature more broadly, we contribute a framework that is informed by subjects from a virtuality context otherwise understudied \cite{Edmondson_Bransby_2023}. Not unlike most PS research \cite{Edmondson_Bransby_2023}, the recent, comprehensive set of PS categories by \citet{O’Donovan_Van_Dun_McAuliffe_2020} (i.e., voice, support, familiarity, and learning) is informed by, and tailored for, healthcare teams. How PS manifests is context-dependent, and our findings suggest the same. For example, the \textit{embrace mistakes} sub-theme was informed by our teams describing software bugs as \textit{inevitable} aspects of development (as opposed to \textit{avoidable errors}). The emphasis on innovation speed and experiential learning thus illustrates how our context differs from one where patient safety instead drives PS \cite{O’Donovan_Mcauliffe_2020_syst}, underscoring the need to ground PS research in the realities of the teams we aim to support.

While contributing a study of PS in hybrid teams, none of our 24 indicators are necessarily constrained to a single hybrid modality---RQ1 results were derived by asking hybrid engineering design teams about PS indicators without specific reference to in-person or Slack interactions, since we know that all team interactions, regardless of hybrid modality, will affect PS. Our indicators may then transfer to PS in innovation teams of various virtuality contexts.
In fact, 
a review of PS literature identifies parallels between our five facets and findings from primarily in-person team contexts \cite{Edmondson_Bransby_2023}: PS is known to promote voice and speaking up behaviours \cite{Edmondson_Bransby_2023, Edmondson_Lei_2014, Frazier_Fainshmidt_Klinger_Pezeshkan_Vracheva_2017} (Communication); drive teamwork, individual performance, and engagement \cite{Edmondson_Bransby_2023, Edmondson_Lei_2014, Frazier_Fainshmidt_Klinger_Pezeshkan_Vracheva_2017} (Cooperation); foster supportive, inclusive, and authentic climates \cite{Edmondson_Bransby_2023} and be fostered by familiarity and support between colleagues \cite{Frazier_Fainshmidt_Klinger_Pezeshkan_Vracheva_2017, O’Donovan_Mcauliffe_2020_syst} (Culture); play a mediating role between leadership behaviours and team outcomes, and be reliant on positive leadership styles, hierarchy, and status \cite{Edmondson_Bransby_2023, Edmondson_Lei_2014, Frazier_Fainshmidt_Klinger_Pezeshkan_Vracheva_2017, O’Donovan_Mcauliffe_2020_syst} (Leadership); facilitate team learning behaviours \cite{Edmondson_Bransby_2023, Edmondson_Lei_2014, Frazier_Fainshmidt_Klinger_Pezeshkan_Vracheva_2017} and be facilitated by learning orientations \cite{Frazier_Fainshmidt_Klinger_Pezeshkan_Vracheva_2017, O’Donovan_Mcauliffe_2020_syst} (Learning). These parallels instill confidence in our five-facet framework, whose novelty and organization should enable leaders of all work contexts to easily monitor for, and nurture, team PS.

Our indicators offer empirical support for, or additional detail in, a number of past PS studies. For example, 
our \textit{gratitude} indicator extends findings that \textit{leaders’} trait gratitude is associated with perceptions of PS \cite{Li_Wu_Brown_Dong_2022}, to suggest \textit{team-wide} gratitude as significant, too. We also extend the association between power/status and PS \cite{Edmondson_Bransby_2023, Edmondson_Lei_2014, Lee_Choi_Kim_2018, Nembhard_Edmondson_2006, O’Donovan_Mcauliffe_2020_syst}, where status conflict is negatively associated with PS \cite{Lee_Choi_Kim_2018}, to suggest that even unwritten hierarchies (e.g., based on the technical rigour of a sub-team’s work, as described by P10) interplay with PS in multidisciplinary teams. 
Our study also illuminates potentially contrasting effects of time pressure on PS, through Learning. Our teams described time pressure as an obstacle to skill learning, but a reason to accept mistakes and obvious questions. While a negative association between time pressure and PS climate has been found \cite{Silla_Gamero_2018}, our qualitative findings suggest a nuanced interplay with learning behaviours, where mistakes are more readily accepted but learning new skills becomes more challenging.

\subsubsection{PS in HCI}
RQ1 additionally contributes qualitative insights to the largely quantitative PS findings in HCI. First regarding Hastings and colleagues' findings \cite{Hastings_Jahanbakhsh_Karahalios_Marinov_Bailey_2018} that PS is positively correlated with measures of team experience, including satisfaction with the team, our interview-informed framework offers rich insights on behaviours and norms that may be associated with these relationships. For example, \textit{dedication} (under Cooperation) described consideration for one's teammates as a positive PS indicator (a form of mutual respect). Such behaviours may then encourage satisfaction with one's team. Our framework can similarly serve as a starting point to explore potential explanations for the findings in \citet{Cao_Yang_Chen_Lee_Stone_Diarrassouba_Whiting_Bernstein_2021} and \citet{Park_Santero_Kaneshiro_Lee_2021}, which identify PS-derived measures as positively associated with team viability and efficacy, respectively. Meanwhile, our indicators related to authenticity (or lack thereof), including \textit{delay and discreetness} (under Communication), and \textit{fear} (under Cooperation), may explain why \citet{Musick_Gilman_Duan_McNeese_Knijnenburg_O’Neill_2023} found the anonymized condition of a teammate information system to result in nearly-significant worse PS levels---a culture of interpersonal risk-taking and mutual respect may thrive best when teammates can get to know each other without implicit (e.g., \textit{delay and discreetness}) or explicit (e.g., \textit{anonymization}, as in \citet{Musick_Gilman_Duan_McNeese_Knijnenburg_O’Neill_2023}) buffers.

Our findings also extend that of \citet{Hastings_Jahanbakhsh_Karahalios_Marinov_Bailey_2018}, which found similar PS levels between teams who engaged in team-focused activities (i.e., two scripted exercises) and those who did not---we identify 24 indicators that potentially contribute to PS' build-up, whose incorporation in future experimental studies may result in significant differences in PS levels. Ultimately, our five-facet PS framework serves not only as actionable support for (hybrid) teams invested in interpersonal risk-taking and mutual respect, but also as a starting point for HCI researchers interested in the specific behaviours that encompass PS and its associations with other team outcomes.

\subsection{Socio-technical Gaps in and Design Implications for Instant-Messaging Platforms}


Through RQ2 results, we contribute to the limited research on the interplay between instant-messaging platforms, work, and social context \cite{Lee_Yuan_2024, gupta2013should}, an understanding of how hybrid engineering design teams perceive four facets of PS to differ in the extent to which they are effectively expressed on Slack compared to in-person. While \citet{Lee_Yuan_2024} recently identified subordinates' impression management and self-censorship tendencies over instant-messaging with supervisors, our \textit{prone to in-authenticity} discovery (under Culture) suggests that such behaviours may affect PS and span teams---not only supervisor-subordinate relationships. There are additional parallels between overly-task focused communication \cite{Lee_Yuan_2024, huang2019engagement, herbsleb2002introducing} and our \textit{purely business} sub-theme; low engagement \cite{Lee_Yuan_2024, huang2019engagement, herbsleb2002introducing} and our \textit{disengagement and inactiveness} sub-theme; misinterpretations \cite{Lee_Yuan_2024, herbsleb2002introducing} and our \textit{ineffective discussions} sub-theme, to which we contribute additional perceived effects alongside an understanding of how they may be explained through a single team construct---PS. We thus reveal technology gaps through the lens of relational challenges, as recommended \cite{Bjorn_Busboom_Duckert_Bødker_Shklovski_Hoggan_Dunn_Mu_Barkhuus_Boulus-Rødje_2024}, for the first time with PS in HCI.

In recognition of PS' significance for team-level outcomes, including innovation, learning, and engagement \cite{Edmondson_Lei_2014, Edmondson_Bransby_2023, Edmondson_1999}, we ground our study in this mature construct \cite{Edmondson_Bransby_2023} to leverage psychology and management literature in the design of collaborative technology (instant-messaging). In turn, we proceed to present three design implications with accompanying examples, to support PS in teams who use instant-messaging: 1) lowering the stakes, 2) regulating turn-taking, and 3) intentionally engaging. While we present the design examples on Slack-like interfaces for illustration purposes, we imagine that these features could be implemented in most instant-messaging platforms.

\subsubsection{Lower the Stakes}
We found that teammates are more muted and careful with Slack communication since texts are devoid of tone and permanently stored. Similarly, critiquing on Slack was inhibited out of fear of hypocritical message histories and the inability to quickly pivot ideas without immediate feedback. Evidently, the permanence of Slack messages hinders participants’ willingness to engage in Psychologically Safe behaviours. In response, our first design example \textit{lowers the stakes} on Slack, encouraging frictionless and authentic dialogue. 

\begin{figure}[h]
\centering
\includegraphics[scale=0.40]{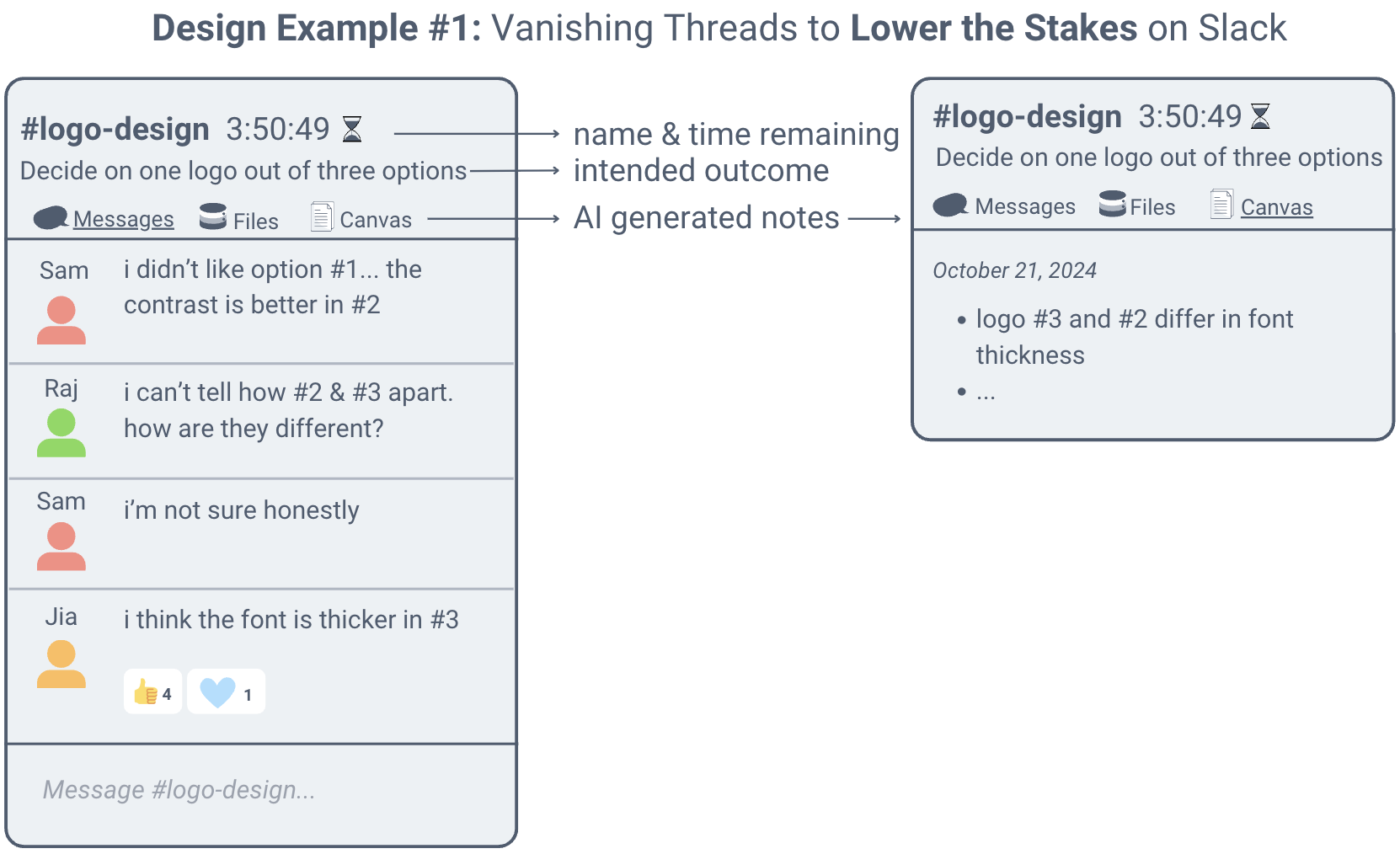}
\Description{To illustrate the \textit{lower the stakes} design implication, we envision vanishing threads on Slack. These threads would include timers to indicate time remaining until all messages are cleared. They would also have associated canvases, where team members/AI would document salient but anonymous points from the cleared discussions, ensuring adequate documentation.}  
\caption{Design Example \#1: We illustrate the \textit{lower the stakes} design implication using vanishing threads as a design example: temporary threads where participants’ identities are visible (left). AI documents salient (but anonymized) notes from the vanishing threads to prevent information loss (right)}
\label{fig:design_example_1}
\end{figure}

We propose Slack threads with a ``vanish'' mode, where messages are temporary (Figure \ref{fig:design_example_1}). To ensure adequate documentation of these vanishing discussions, there can be associated shared canvases (Slack-based collaborative documents) where teammates (or AI) can note salient outcomes without any user identifiers. Anyone can launch a vanishing thread request, encouraging all teammates to initiate free-flowing discussions. Threads must be accompanied by an intended outcome statement, e.g., ``decide on the logo'', ensuring that their purpose is clear.
We envision vanishing threads as being useful for spontaneous ideation and feedback sessions. We considered that participant names could be anonymized in these threads, in line with literature that suggests anonymity as useful for upward communication \cite{Zhan_Wan_Sun_2022}, creativity \cite{Baruah_Green_2023}, team satisfaction and production of high-quality ideas \cite{Pissarra_Jesuino_2005}, and curtailing the negative effect(s) of status differences on performance \cite{Chang_2011}. However, we opted out of such a design decision, first for Musick and colleagues' \cite{Musick_Gilman_Duan_McNeese_Knijnenburg_O’Neill_2023} recent finding that anonymity in system design may significantly lower team satisfaction and team cohesion levels \cite{Musick_Gilman_Duan_McNeese_Knijnenburg_O’Neill_2023}. Second---our own participants expressed concerns about inauthenticity and fear among teammates, which we worry may be further normalized by anonymization. As such, we consider identifiable (but temporary) threads that are accompanied by anonymized documentation as a practical middle ground. Vanishing threads also align with Liu and colleagues' \cite{Liu_Van_Essen_Eggen_2024} recent design space schema for more \textit{human-like} experiences in technology for hybrid teams, specifically their mentions of information delivery through \textit{ambient design} and \textit{light-weight interactions}. These insights further motivate \textit{lower the stakes} as a design implication for greater PS support.
\subsubsection{Regulate Turn-taking}
From the Learning facet of PS, we saw that Slack discussions are occasionally ineffective, with a tendency to turn messy when multiple people are chiming in at a time. As a result, teams would move and delay discussions to in-person settings. Acknowledging the need for productive and organized Slack discussions, our next design implication prioritizes orderly Slack participation.

\begin{figure}[h]
\centering
\includegraphics[scale=0.38]{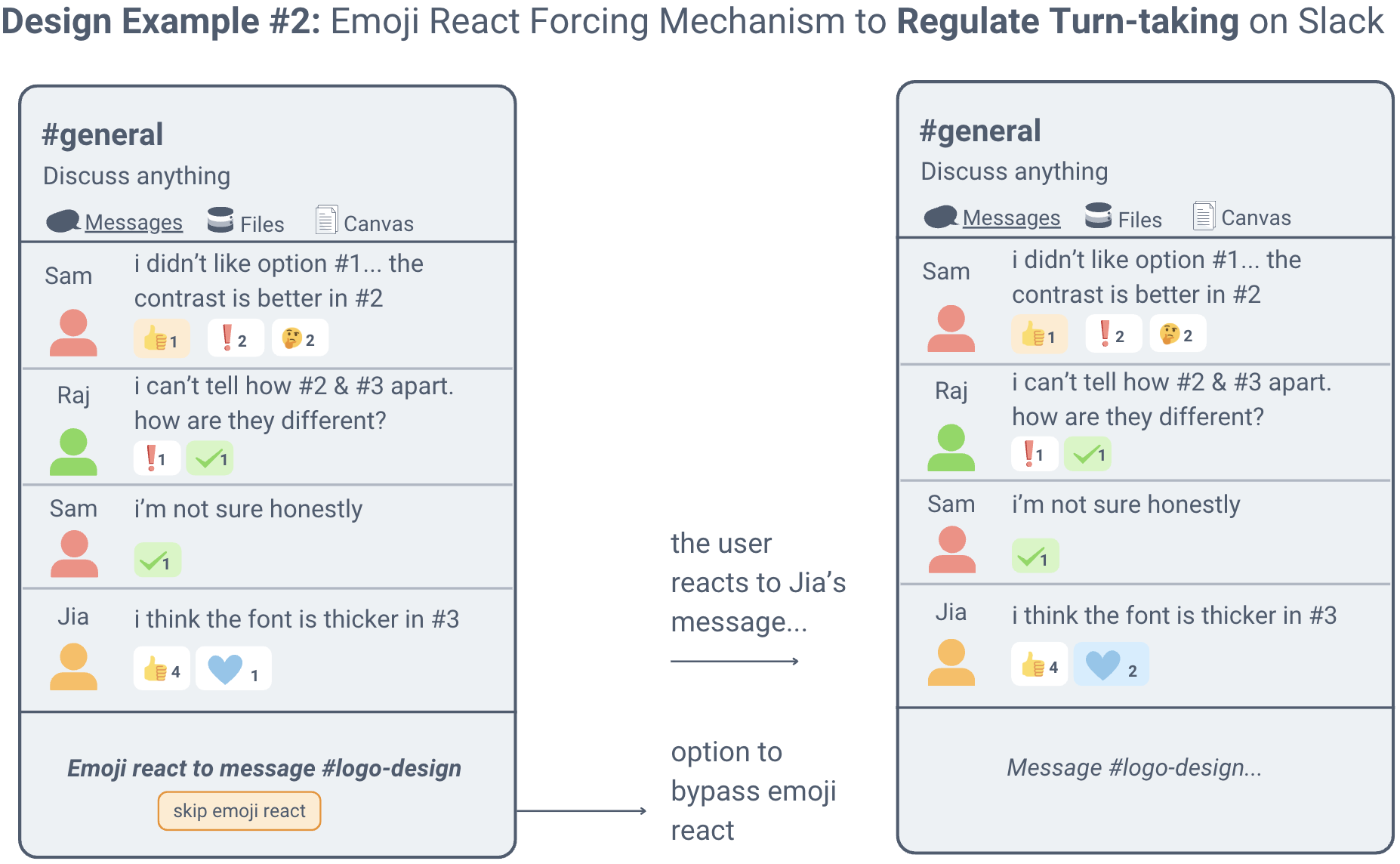}
\Description{To illustrate the \textit{regulate turn-taking} design implication, we imagine emoji react forcing mechanisms. The mechanism, when turned on, would require a teammate to emoji react to all of their teammates' messages before sending a message to the same location (e.g., a Slack channel). Users of the mechanism can easily bypass the mechanism when it is turned on by pressing a button that says 'skip emoji react.'}  
\caption{Design Example \#2: We illustrate the \textit{regulate turn-taking} design implication using emoji react-forcing mechanisms as a design example. When users enable the mechanism in a conversation, they are prevented from sending a new message until they emoji react to all their teammates’ messages or bypass the mechanism (left). A user regains the ability to send messages if they emoji react to all their teammates’ messages (right).}
\label{fig:design_example_2}
\end{figure}

We envision a design that prevents users from sending messages in channels with three or more individuals until they emoji react (acknowledge) all previous messages sent by their teammates in the same location (Figure \ref{fig:design_example_2}). To prevent users from aimlessly emoji reacting without having read the message(s), the option to emoji react will only become available after set durations, depending on the user's reading speed. The forced delays prevent multiple teammates from sending messages simultaneously, thereby enhancing chat readability.

We emphasize orderly participation because turn-taking---a characteristic of high PS teams \cite{Duhigg_2016}---is a strategy to foster PS \cite{Kolbe_Eppich_Rudolph_Meguerdichian_Catena_Cripps_Grant_Cheng_2020} and associated with teams’ higher perceptions of PS \cite{O’Connell_Cole_Mohammed_Jablokow_Miller_2022}. HCI researchers have found eye gaze patterns to facilitate turn-taking \cite{Jokinen_Furukawa_Nishida_Yamamoto_2013}. As such, CSCW researchers have built tools that provide sociometric feedback to
balance participation in teams who communicate verbally \cite{DiMicco_Pandolfo_Bender_2004, Kim_Chang_Holland_Pentland_2008}. This design proposes similar support for instant-messaging, 
such that users are encouraged to pause and digest team messages before contributing new ideas. 

Forced delays between messages may slow progress during discussions and obstruct time-sensitive announcements. Accordingly, the turn-taking mechanism would be off by default and always be optional, preserving user autonomy. Only when a user is participating in a discussion with more than two teammates would they be prompted to use the turn-taking mechanism for a temporary duration. 
There is also the concern that forced turn-taking may force disingenuous engagement with teammates’ messages. While alternative reaction mechanisms---such as ReactionBot in CSCW \cite{Liu_Wong_Pudipeddi_Hou_Wang_Hsieh_2018}---exist to encourage genuine interactions on Slack, they may also inspire anxieties about emotional leakage and self-presentation management \cite{Liu_Wong_Pudipeddi_Hou_Wang_Hsieh_2018}. Thus, our mechanism would recommend the use of neutral emojis (e.g., checkmark) when an emoji that implies a sentiment may be unsuitable or disingenuous. These properties of the turn-taking mechanism ensure that users remain accountable for reading messages and honest in their reactions. 

\subsubsection{Intentional Engagement}
We additionally learned about teammates who are inactive, and less willing to respond to requests for feedback and help, on Slack. This is frustrating for the teammates whose requests are met with little to no engagement. Acknowledging that teammates have varying and unmet expectations about Slack participation, our final design implication prioritizes the intentional use of online statuses.
\begin{figure}[h]
\centering
\includegraphics[scale=0.53]{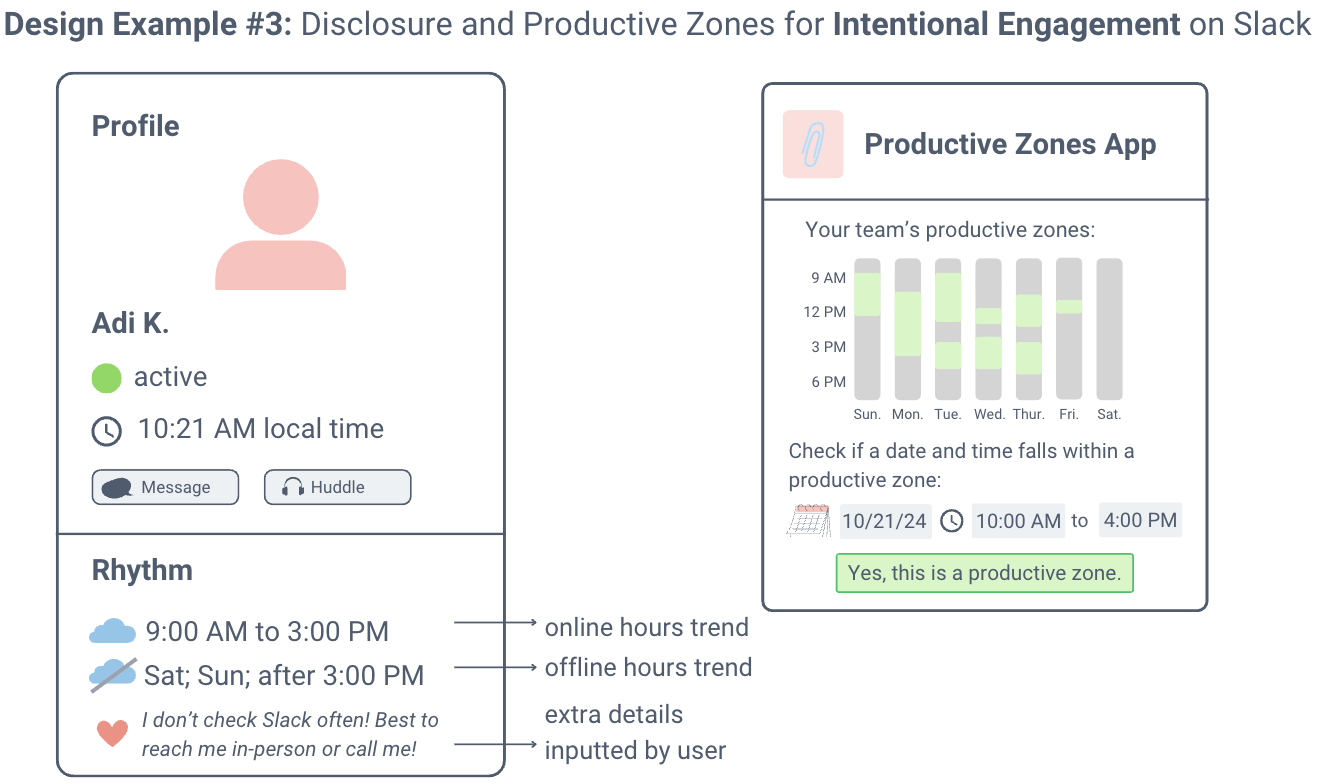}
\Description{To illustrate \textit{intentional engagement}, we use a design example that consists of informative online statuses on Slack profiles, where a teammates' usual online hours, usual offline hours, and work preferences (e.g., “I don't check Slack often! Best to ask me questions in-person or call me!”). An AI-powered productive zone application may also be useful, where teammates can see when most of their teammates tend to be online, and use this information to optimally select times to initiate discussions on Slack.}  
\caption{Design Example \#3: We illustrate the \textit{intentional engagement} design implication using work rhythm disclosure (left) to support teammates while they navigate communication decisions. The productive zones application (right) is an AI-powered tool that indicates when most teammates are available on Slack.}
\label{fig:design_example_3}
\end{figure}
We encourage teammates to keep online statuses up-to-date and disclose their work preferences (e.g., “I prefer questions in-person”) on their profiles. With everyone aware of each other's work rhythms, teammates can be strategic in who they contact, when, and where. For example, an individual on the Mechanical team with a question for the Design team can target Design teammates who are likely online, instead of offline members or the whole team (prompting the bystander effect). We also envision automated analysis of team activity to define team-wide productive zones (i.e., periods), inspired by virtual teams who leverage these periods already  \cite{Breideband_Talkad_Sukumar_Mark_Caruso_D’Mello_Striegel_2022}. Leaders can then optimally select when to initiate discussions for maximal participation, no longer delaying discussions to in-person meetings. See Figure \ref{fig:design_example_3} for a mock-up.

This design is inspired by HCI research that emphasizes digital footprints and social awareness \cite{Yang_Yamashita_Kuzuoka_Wang_Foong_2022}, including the intentional use of online status \cite{Breideband_Talkad_Sukumar_Mark_Caruso_D’Mello_Striegel_2022}. Such awareness cues nurture communication decisions, communication expectations, and relational connectedness \cite{chou2022because}. Transparency about, and accommodation for, diverse work rhythms also allow teams to thrive with work-rhythm diversity
\cite{Breideband_Talkad_Sukumar_Mark_Caruso_D’Mello_Striegel_2022}. 
Moreover, \citet{chou2022because} finds that individuals are willing to sacrifice their privacy through status indicators to justify communication boundaries. 
Our design implication nonetheless does not address inflexible teammates who will remain inactive on Slack. Instead, it promotes greater awareness in teamwork and communication patterns, preventing teammates from being \textit{incorrectly perceived} as uninvolved, and encouraging optimal team communication decisions.

\subsection{Future Automation of PS Measurement}
\label{sec:discussion-slack-signals}
Through RQ3, we contributed 15 Slack-based PS indicators whose deployment across six hybrid teams resulted in observable, unique trends between teams and indicators across a 12-week timeline. We thus respond to the need to observe how PS evolves in teams over time \cite{Edmondson_Bransby_2023, Edmondson_Lei_2014}, through design of indicators that show potential in revealing longitudinal trends. Above all, our Slack-based indicators establish the foundation for online PS measurement on instant-messaging platforms, in a field where measures have yet to capture behavioural dynamics in nontraditional settings \cite{Edmondson_Bransby_2023}.

Whereas recent HCI work has explored PS in messages \cite{Cao_Yang_Chen_Lee_Stone_Diarrassouba_Whiting_Bernstein_2021, Ferguson_Van_De_Zande_Olechowski_2024}, our study is the first to leverage hybrid teams' perceptions of how PS manifests to inform the design of Slack-based PS indicators. This first contrasts Cao and colleagues' \cite{Cao_Yang_Chen_Lee_Stone_Diarrassouba_Whiting_Bernstein_2021} reliance on human coders for manual and ordinal PS labels in their study of team viability predictors in text messages---our 15 indicators enable purely quantitative investigations, which are more scalable and reproducible than traditional qualitative approaches \cite{antons2020application}. Our work additionally extends \citet{Ferguson_Van_De_Zande_Olechowski_2024}---exploratory work that proposes emojis, keywords, and actions (e.g., replies) as potentially differentiating low versus high PS teams. Different from Ferguson and colleagues' \cite{Ferguson_Van_De_Zande_Olechowski_2024} translation of \citet{O’Donovan_Van_Dun_McAuliffe_2020}'s PS categories (as informed by healthcare teams) to quantitative PS indicators in text messages, we use interviews with hybrid innovation teams to inform our PS framework and indicators, grounded in Edmondson's \cite{Edmondson_1999} PS operationalization. As a result, our Slack-based indicators encompass behaviours more relevant to innovation contexts (e.g., \textit{embrace mistakes}), as discussed in Section \ref{discuss:actionable}. Having observed trends that differentiate teams between and within the PS facets in Section \ref{res:slack-trends-heatmaps}, we are additionally confident that our indicators lay the groundwork for future efforts toward an automated PS measure. Naturally, a team's PS is ultimately signalled and affected by all team interactions---both in-person and online. Herein, we contribute measurement support that will enable leaders to monitor for, and nurture, online PS on Slack, to ensure that PS is as recognizable and prioritized online as it is in-person.

\section{Limitations \& Future Work}
\label{sec:limits}

We cannot claim that our findings generalize to all hybrid teams since our subjects were sourced from one engineering design course context. Our sample's uniformity had its strengths, namely in the confidence that all participants came from teams with similar Slack and in-person collaboration norms, as mandated by course resources and expectations. We additionally documented parallels between our findings and those of broader PS literature. Still, our study of student teams, like other recent HCI work \cite{ferguson2023we, Musick_Gilman_Duan_McNeese_Knijnenburg_O’Neill_2023}, means that contextual factors, such as academic (not financial) objectives, could have resulted in PS differences from industry teams.  We look forward to studying more representative innovation teams and additional collaboration platforms in the future.

We additionally caution readers from considering our Slack-based PS indicators as validated measures. On the contrary, the indicators serve as a proof of concept for now---we used RQ3 to demonstrate them as capable of unveiling PS differences between teams, facets, and indicators. We look forward to validating and refining the indicators' design through comparison with traditional surveys, and interview and observational studies, such that they may be used by leaders interested in nurturing team PS over instant-messaging platforms. Regarding surveys, we will leverage the Experience Sampling Method \cite{larson2014experience} for frequent and repeat PS reports by each teammate. This will ensure that we collect longitudinal, high-volume validation data for fair and meaningful comparison with the continuous data points derived by Slack-based analysis. Our survey questions will specify whether responses should refer to in-person, Slack-based, or hybrid team interactions, as motivated by our RQ2 findings. We will also investigate and ensure the Slack-based PS indicators' discriminant validity prior to averaging all metrics to create a single score, with particular attention to components that currently contribute to multiple indicators (e.g., first-person plural pronouns).

Importantly, the teams that we studied were diverse in their gender and racial/ethnic composition. While our interview sample of PMs and HRMs also comprised a gender and racial/ethnic mix, it was skewed towards those who identify as men, potentially due to the between-team variance in gender diversity among leaders. Our pursuit of leader-only interviewees was intentional, but the resulting gender imbalance in our interview sample could have biased the development of the Slack-based PS indicators. As such, the indicators will be validated against a larger sample of teams of various demographic compositions in future work.


We must also clarify that a team's PS, regardless of where it is measured, is likely affected by the team's interactions \textit{everywhere}. It is imperative that users of future automated PS measures approach PS holistically, and consider both a mix of observable behaviours (our five-facet PS framework) and quantitative measures across hybrid modalities (e.g., in-person, Slack) in gauging the health of a team's interpersonal engagements. The need for holistic PS measurement is further stressed by our RQ2 results, wherein we found that PS \textit{on Slack} is perceived to be different from \textit{in-person} PS. As such, these indicators cannot be classified as ``complete'' PS measures---not unless we reach a point of confidence in PS being perfectly preserved across hybrid modalities. Instead, the Slack-based PS indicators serve as a critical sub-component of future ``complete'' PS measures that honour the need to support teams with nurturing and monitoring online PS as it presents in an instant-messaging platform environment. We encourage future work to inform and design the other critical pieces (e.g., PS monitoring for video-conferencing), all in pursuit of an eventual ``complete'' PS measure that supports hybrid teams' PS \textit{everywhere}.


\section{Conclusion}
To nurture interpersonal trust and mutual respect in hybrid teams, we used interviews to study six engineering design capstone teams' perceptions of PS and how it differs on Slack from in-person. We found that PS presents in their Communication, Cooperation, Culture, Learning, and Leadership and Status. Participants also perceived indicators from four PS facets to differ in the extent to which they are effectively expressed on Slack compared to in-person interactions---\textit{suggestive of a second side to PS}. We contribute 24 PS indicators, three design examples to support PS on Slack, and 15 Slack-based PS indicators to monitor and nurture hybrid teams' interpersonal engagements across hybrid modalities. Having also leveraged teams' instant-messaging data, we illustrate how HCI researchers may use modern data sources to explore how team-level constructs present differently across hybrid modalities---potentially finding that there are (at least) \textit{two sides to every story}.

\section{Acknowledgements}
We are grateful to Dr. Georgia Van de Zande (Olin College of Engineering and Massachusetts Institute of Technology) and the teaching team for their contributions to our collection and understanding of the data used in this work. We also thank the associate editor and reviewers for their valuable feedback.

\bibliographystyle{ACM-Reference-Format}
\bibliography{sample-base}


\newpage
\appendix
\section{Codebooks} \label{appendix:codebook}
\subsection{RQ1 Codebook with Example Codes}
\begin{table}[h]
\footnotesize
\caption{Final Codebook with Example Codes for RQ1. The starred sub-themes are the negative PS indicators.}
\begin{tabular}{p{0.13\linewidth}|p{0.3\linewidth}|p{0.5\linewidth}}
\hline
Theme     & Sub-theme& Example\\ \hline
Communication   & Collective decision-making & \textit{Any time that decisions are being made… it's generally a conversation.}\\
    & *Delay and discreetness    & \textit{...They didn't tell us until a few days later.}  \\
    & *Dominant voices     & \textit{The rest of the team isn't as comfortable to just say it.} \\
    & Respectful and inclusive language  & \textit{Our team uses [an internal phase that reminds teammates to ground their ideas in facts] a lot, and it's very effective. It just deescalates a situation very quickly.}\\
\textbf{} & Turn-taking    & \textit{It's like, `Oh, hey! I think it's your turn to talk', and like giving them that opportunity.}  \\ \hline
Cooperation     & *Stubborn& \textit{They ended up stepping back from the team a little bit...They felt unheard because we didn't go with their idea.}   \\
    & Dedication     & \textit{People were pretty receptive...`Yeah, we realized that you guys seem pretty spread thin, a little frazzled. We want to step up.'} \\
    & *Low effort    & \textit{They show up with absolutely nothing {[}done{]}.}    \\
    & Initiative     & \textit{Our team is very driven and wants to be able to do things well…There's like almost a little bit of pride… It's not… ‘Can I pawn it off to somebody else?’}  \\ \hline
Culture   & Emphasis on community-building& \textit{Generally, there's like a positive dynamic on the team…everyone wants everyone to feel like that part of the team.}    \\
    & *Fear    & \textit{They don't want to speak directly to the person cause, they don't want to seem uncool.}  \\
    & Expressions of gratitude& \textit{Any member who's willing to commit to the team and put in time and effort… everybody on the team respects that so much.}\\
    & Accountability without blame & \textit{Playing the blame game isn't helpful.}   \\ \hline
Leadership and status & *Disengaged leaders  & \textit{Our [PM]... will go, `Everything went great. You guys are perfect...' And then I go, `No...here's a list of like 30 problems.' } \\
    & *Overwhelmed leaders & \textit{We had this dynamic within the team where the {[}PMs{]} would put in a lot more hours than other members.} \\
    & Shared leadership    & \textit{I think people having ownership over what they’re doing helps them speak about it.}\\
    & *Social hierarchy    & \textit{They think it'll be more successful if I champion it...}    \\ \hline
Learning  & Adapt and improve    & \textit{They took it pretty well. We seem to have moved forward...}  \\
    & Embrace mistakes     & \textit{No one's going to get upset when that inevitable situation occurs.}    \\
    & Embrace obvious questions  & \textit{People can ask questions that you may deem stupid.}  \\
    & Embrace critiques    & \textit{Everybody on our team can appreciate getting called out when they're starting to get like nonfactual.}     \\
    & Embrace new ideas    & \textit{It could be something in our back pocket… We aren't doing this for this iteration, {[}but{]} maybe for future iterations.}   \\
    & Seek help& \textit{If somebody doesn't know how to do something, they just say so.}     \\
    & Teach    & \textit{They've been really awesome at teaching them.} \\ \hline     
\end{tabular}
\end{table}

\newpage
\subsection{RQ2 Codebook with Example Codes}
\begin{table}[H]
\footnotesize
\caption{Final Codebook with Example Codes for RQ2.}
\label{table_appendix:codebook}
\begin{tabular}{l|p{0.3\linewidth}|p{0.5\linewidth}}
\hline
Theme   & Sub-theme     & Example  \\ \hline
Communication  & Cautiousness  & \textit{It has a paper trail... I don't want there to be a   paper trail of anything but like could be interpreted negatively...}\\
  & Unclear and   inefficient & \textit{You may spell something wrong. You may misphrase   something…}\\ \hline
Cooperation   & Disagreeable teammates & \textit{I think it's like a tendency to be like   100\% attached your idea, and like not willing to compromise on Slack.}  \\
\textbf{}     & Untrusting of teammates   & \textit{On Slack…I find myself thinking, `Oh, are they upset?'   … `Oh, do they think I'm stupid?'}\\
\textbf{}     & Disengagement and inactiveness    & \textit{I think people   just don't read Slack all that much.} \\ \hline
Culture & Prone to inauthenticity   & \textit{I didn't see any super negative reactions [to that message], but also I don't know if people would react negatively on Slack...} \\
  & Purely business     & \textit{If you're going   {[}to{]} admit something on Slack… I feel like it's more related to the schedule.}   \\ \hline
Learning& Mistakes are easy to miss & \textit{I was like kind of a day late {[}to   giving feedback on the schematic{]} … Other people hadn't caught that {[}mistake{]}   yet.}    \\
  & Feedback is weak and delayed to in-person    & \textit{Once brought up {[}in-person{]}… {[}It{]} was like, `Oh, wait!   Can we all add to this a little bit?...'}   \\
  & Ineffective discussions that are prone to misunderstandings    & \textit{It's so much  easier to de-escalate and see like their tone and those kinds of things in-person.}     \\ \hline
\end{tabular}
\end{table}

\end{document}